\documentclass[11pt]{amsart}

\usepackage[english]{babel}
\usepackage[left=3.2cm,right=3.2cm,top=2.9cm,bottom=2.9cm]{geometry}
\usepackage{color,enumitem,graphicx}
\usepackage{amsmath}
\usepackage{graphicx}
\usepackage{amssymb}
\usepackage{subfigure}
\usepackage{setspace}
\usepackage{epic}
\usepackage{bbold}
\usepackage{color}

\newtheorem{theorem}{Theorem}[section]

\input epsf
\usepackage{url,mathrsfs}

\newcommand{\cB}{\mathcal{B}}

\newcommand{\f}{\mathbf{f}}

\newcommand{\sm}{\smallskip}

\newcommand{\no}{\noindent}

\newcommand{\RR}{\mathbb{R}}

\newcommand{\E}{\mathbb{E}}

\def\no{\noindent}

\def\sm{\smallskip}

\def\va{\raise 2pt\hbox{,}}

\def\boxitt#1{\vbox{\hrule\hbox{\vrule\kern3pt
\vbox{\kern7pt#1\kern7pt}\kern3pt\vrule}\hrule}}

\numberwithin{equation}{section}

\title[Modeling the Onset and Evolution of Criminality]
      {FROM A SYSTEMS THEORY OF SOCIOLOGY TO MODELING THE ONSET AND EVOLUTION OF CRIMINALITY}

\author[N. Bellomo, F. Colasuonno, D. Knopoff, J.Soler]{}


\thanks{The first part of this paper, devoted to methodological topics,  received funding from the European Union's Seventh Framework Programme (FP7/2007--2013) under grant agreement  (SAFECITI).\\ Project full title: ``Simulation Platform for the Analysis of Crowd Turmoil in Urban Environments with Training and Predictive Capabilities''. The application developed in the second part does not specifically refer to the project. This publication reflects the views only of the author, and the Commission cannot be held responsible for any use which may be made of the information contained therein. The paper has been also partially supported by Junta de Andaluc\' ia Project FQM 954, while the fourth author has been also partially supported by MINECO (Spain).}

\begin{document}
\maketitle

%% Enter the first author's name and address:
\centerline{\scshape Nicola Bellomo}
\medskip

{\footnotesize
 %% please put the address of the first author
 \centerline{Department of Mathematics, Faculty  Sciences, King Abdulaziz University,}
   \centerline{Jeddah, Saudi Arabia.}
      \centerline{Politecnico Torino, Corso Duca degli Abruzzi 24, 10129 Torino, Italy}
}
\medskip

\centerline{\scshape Francesca Colasuonno}
\medskip

{\footnotesize
 %% please put the address of the first author
 \centerline{Department of Mathematical Sciences, Politecnico of Torino,
Corso Duca degli Abruzzi 24,}
   \centerline{10129, Torino, Italy}
} %% Do not forget to end the {\footnotesize by the sign }
\medskip

\centerline{\scshape Dami\'an Knopoff}
\medskip

{\footnotesize
 %% please put the address of the first author
 \centerline{Centro de Investigaci\'on y Estudios de Matem\'atica (CONICET),\\}
 \centerline{Medina Allende s/n, 5000 C\'ordoba, Argentina}
}
  %% Do not forget to end the {\footnotesize by the sign }
\medskip

\centerline{\scshape Juan Soler}
\medskip

{\footnotesize
 %% please put the address of the first author
 \centerline{Department of Mathematics, University of Granada, Spain}
} %% Do not forget to end the {\footnotesize by the sign }

\medskip

\begin{abstract}
This paper proposes a systems theory approach  to the modeling of onset and evolution of criminality in a territory, which  aims at capturing the complexity features of  social systems. Complexity is related to the fact that individuals  have the ability to develop specific heterogeneously distributed strategies, which depend also on those expressed by the other individuals. The modeling is developed by methods of generalized kinetic theory  where interactions and decisional processes are modeled by theoretical tools of stochastic game theory.

\end{abstract}

\parindent=20pt

%%%%%%%%%%%%%%%%%%%%%%%%%%%%%%%%%%%%%%%%%%%%%%%
%%%%%%%%%%%%%%%%%%%%%%%%%%%%%%%%%%%%%%%%%%%%%%%

\section{Plan of the Paper}

This paper aims at developing a  systems theory approach  to the modeling  of criminality phenomena in a territory.   This objective is pursued bearing in mind the idea that all living systems in general, and social systems in particular,  are complex \cite{[BAI94]}. This feature, as observed in \cite{[BHT13]}, as well as in the collection of articles \cite{[BALL12]}, is related to the fact that individuals  have the ability to develop specific strategies, which depend also on those expressed by the other individuals. These strategies are heterogeneously distributed. Moreover, individuals in socio-economic systems are able to learn from their experience. This implies that the expression of the strategy evolves in time, and consequently that interaction dynamics undergo modifications.

The main feature of these systems is that their collective overall behavior is determined by the dynamics of their interactions, while the  modeling of individual dynamics does not lead in a straightforward way to a mathematical description of collective emerging behaviors. Therefore, the challenging objective of the modeling  consists in transferring  the dynamics at the scale of individual entities into collective emerging behaviors.

The contents are mainly motivated by two well defined hints. The first one is related to the idea that the dynamical processes in Economics and Social Sciences are highly affected by individual (rational or irrational) behaviors, reactions, and interactions. These concepts  have begun to impose themselves  to the traditional assumption of rational socio-economic behavior, starting from the concept of bounded rationality \cite{[SIM97]}. Therefore, the contribution of mathematics to a deeper understanding of the relationships between individual behaviors and collective outcomes may be fundamental \cite{[BHT13]}. This new methodological approach looks at Economics, and Social Sciences in general, as evolving complex systems, where interactions among heterogeneous individuals can even produce unpredictable emerging outcomes \cite{[ADL97],[KV00]}. In this context,  mathematical tools are required to capture the evolving features of socio-economic systems and  incorporate some of their main complexity features. Possibly up to the ability to predict  the so-called \emph{black swan}, which is defined to be a rare event, showing up as an irrational collective trend generated by possibly rational individual behaviors \cite{[TAL07]}.

This paper focuses on criminality being motivated by an increasing interest in the public safety and security in cities. An important contribution is that of Felson \cite{[Fe10]} who gives -from a criminologist's point of view- some basic hints for mathematicians concerning the modeling of crime. He introduces the presence of three fundamental elements that participate in the phenomena, namely offenders, targets and guardians.
Additional phenomenological interpretations are given in \cite{[HHM10]} and \cite{[ORM05]}.

Many of the existing models refer to the spatially homogeneous case, while some others introduce the geographic or spatially dependent component in  kinetic theory \cite{[MS12]} and statistical \cite{[DFWB13],[SDPTBC08]} methods. Tools of games theory \cite{[NOW06]} have been successfully used to model criminality, starting from the so-called prisoner's dilemma and moving towards more sophisticated games. In \cite{[SBD10]}, a game involving $N$ players is considered, where each of them may express four different strategies: paladin, apathetic, informant or villain, while in \cite{[MSB13]} a network study is considered by taking into account subgroups of players through sacred value networks \cite{[TET03]}. Methodologically different approaches can be based on deterministic dynamical systems \cite{[BCG14],[NHP11]}.

The second hint is offered by recent developments of methods of kinetic theory and statistical mechanics in fields far from the classical one of molecular fluid dynamics. Methods of the  generalized kinetic theory have been first applied to model the social dynamics of insects \cite{[JS92]}, followed by a blow up in a rapidly growing literature covering a variety of applications. Among them and without claim of completeness, modeling of social systems \cite{[ABE08],[DL14]}, opinion formation with dynamics over networks \cite{[KNOP14]}, migration phenomena \cite{[KNOP13]}, selective mutations in epidemiology \cite{[DDS11]} and Darwinian mutation and selection on cancer phenomena contrasted by immune cells \cite{[BF06],[BDK13],[DEA14]}.

The mathematical approach developed in the previously cited papers belongs to the so-called kinetic theory for active particles, for short KTAP approach,  which refers to large population of living entities interacting with rules modeled by theoretical tools of game theory. The collection of surveys \cite{[BP00]} witnesses the beginning of a systematic use of methods of the kinetic theory in a variety of  applications, which  have been subsequently developed in the last decades by several authors, among others \cite{[DMP12],[HEL10],[PT13]}.

Focusing on the contents of the paper, let us consider a population of individuals  distributed in a certain territory, they can be subdivided into citizens, criminals and detectives. Each aggregation of individuals expresses with a certain intensity its own function-strategy. For instance citizens  operate to improve their own wealth, criminals to subtract it to them and to hide themselves from the chase of detectives, who operate to contrast criminality. The modeling approach aims at studying the interplay between individuals of different aggregations and their growth or decay, which can include transition from one population to the other. Various concomitant causes can, in general, play a role in the development or depletion of criminality, for instance migration phenomena and  welfare policy. An interesting objective consists in looking for extreme events, namely the so-called black swan \cite{[TAL07]}.

This paper is organized through five more sections. In detail, Section 2 presents  a phenomenological description of the complex system treated in this paper, subsequently a modeling strategy is proposed. Section 3 introduces, starting from the description of interactions at the scale of individuals, the mathematical structures suitable to provide the basis for the derivation of specific models. A mathematical model is derived in Section 4 based on the said general framework for a society with wealth distribution constant in time.  Some simulations are presented in Section 5 in order to test the predictive ability of the model. Finally, Section 6 focuses on the development of a general  systems theory approach to sociology.

\section{From a Phenomenological Description to a Modeling Strategy}

Let us consider a large population of individuals with a spatially homogeneous distribution over certain territory, say a town or a region.
The specific features of the population are heterogeneously distributed among individuals, who can, however, be subdivided into
different groups such as citizens owning a certain amount of wealth, small or large, criminals of different levels of villainy, and detectives who are in charge of chasing criminals. Moreover, migration phenomena can be possibly taken into account considering that large towns can attract people from surrounding areas.

Let us briefly summarize the approach, proposed in \cite{[BKS13]}, to model complex systems by the KTAP approach.

\begin{itemize}

\item Individuals are viewed as {\em active particles}, that have the ability to express a specific strategy, called {\em activity}, which
defines their micro-state, namely the state at the microscopic scale.

\item Active particles are subdivided into {\em functional subsystems}, such that they express a specific activity for each subsystem.

\item The activity variable is heterogeneously distributed over the particles, while the overall {\em state} of the population of active particles is delivered, for each subsystem, by a distribution function over the micro-state.

\item Interactions between particles are modeled by theoretical tools of game theory, while the equations describing the dynamics of particles are obtained by a balance in the elementary volume in the space of the micro-states. The inflow and outflow of particles, into and from the said volume, is determined by interactions.

\item The solution of mathematical problems, typically initial and initial-boundary value problems, provides the time evolution of the aforesaid distribution function and hence of the macroscopic description obtained by weighted averaged quantities. Emerging behaviors can be depicted by both distribution function and macro-quantities.

\end{itemize}

An approach, according to the specific features  of the class of systems under consideration, can be proposed as described by the following sequential steps:

\begin{enumerate}

\item Subdivision of the overall system into functional subsystems, each of them with the ability of expressing a different activity (strategy);

\item Modeling interactions at the micro-scale, namely between active particles of the same or different functional subsystems;
including  learning dynamics, which might generate transition from one subsystem to the other;

\item  Derivation of a mathematical structure suitable to describe the evolution in time of the distribution function over the micro-state of particles of each subsystem;

\item Derivation of mathematical models related to the structures introduced in the previous item;

\item Analysis and validation of models.

\end{enumerate}

This approach will be formalized, in the next  sections, being aware that it is limited to a simple picture of the complex variety of social and economic dynamics that effectively occur in our society. Nevertheless, we claim that it has to be regarded as a first step towards a more general  systems theory approach to sociology. This preliminary step will  be followed by the study of more complex case studies outlined  in the last section, which looks ahead to research
perspectives.

%%%%%%%%%%%%%%%%%%%%%%%%%%%%%%%%%%%%%%%%%%%%%%%%%%%%%
%%%%%%%%%%%%%%%%%%%%%%%%%%%%%%%%%%%%%%%%%%%%%%%%%%%%%
%%%%%%%%%%%%%%%%%%%%%%%%%%%%%%%%%%%%%%%%%%%%%%%%%%%%%

\section{From Interactions to a Structure Modeling Collective Dynamics}

This section shows how the strategy presented in Section 2 can be transferred into the derivation of a mathematical structure that can offer the framework  suitable to model the collective dynamics involving different categories of citizens  interacting in a territory. The contents are presented through the next three subsections corresponding to the first  sequential steps of the aforesaid strategy, where the last  paragraph also proposes a critical analysis in view of the modeling approach. The approach is based on the assumption that the total number of citizens and the wealth distribution are constant in time.

\subsection{Subdivision into functional subsystems and representation}
Let us consider a population homogeneously distributed in a territory. Individuals in the population are regarded as {\sl active particles} subdivided into a small number of groups according to the specific functions they express in the competition treated in this paper.
The present approach aims at studying how the size of these groups evolves in time, how crime arises and can be controlled in a society, and how the  distribution of the level of different expressions evolves in time.  Therefore, in consonance with the phenomenological description given in the preceding section, the following subdivision into {\sl functional subsystems} is proposed:

\sm \no $i =1$ Normal citizens, whose microscopic state is identified by their wealth, which constitutes the attraction for the eventual perpetration of criminal acts.

\sm \no $i =2$ Criminals, whose microscopic state is given by their criminal ability, namely their ability to succeed in the perpetration of illegal acts.

\sm \no $i =3$ Detectives who chase criminals according to their individual ability.

\sm  The microscopic variables are assumed to be, for each functional subsystem, a real variable $u$ taking values in the subsets $D_1, D_2, D_3 \subset\mathbb{R}^+_0$, respectively. Assuming that a maximal activity value can be identified in each functional subsystem, these three sets are taken, for the sake of simplicity, to be  the interval $[0,1]$, although it is important to keep in mind that the meaning of the variable  differs from one functional subsystem to another. However, worse conditions correspond to lower values of the activity variable, while increasing values of $u$ correspond to higher abilities to express the strategy.

\sm The following table specifies the activity, micro-scale, variable within each functional subsystem:
\begin{table}[h!] \label{table}
\begin{center}
\begin{tabular}{ll}
\hline
\textbf{Functional subsystem} & \textbf{Micro-state} \\
\hline
 & \\						
$i=1$, citizens &	$u\in D_1$, wealth \\
 & \\						
\hline
 & \\						
 $i=2$, criminals &	$u\in D_2$, criminal ability \\	
 & \\						
\hline
 &  \\
 $i=3$, detectives & $u\in D_3$, experience/prestige  \\
 &  \\					
\hline
\end{tabular}
\end{center}
\vskip.2cm
\caption{Microscopic variable for each functional subsystem.}
\end{table}

The representation of the system is delivered by the distribution functions
\begin{equation}\label{distributions}
f_i:[0,T) \times D_i \to \RR_0^+, \qquad i  = 1, 2, 3,
\end{equation}
where $T>0$ is a certain final time, possibly $\infty$. In this way, $f_i(t, u)\, du$ denotes, under suitable integrability conditions, the number of active particles of the functional subsystem $i$ whose state, at time $t$, is in the interval $[u, u + du]$. Therefore
\begin{equation}\label{size}
n_i(t) = \int_{D_i} f_i(t, u)\, du, \quad i = 1, 2, 3,
\end{equation}
defines the {\em size} of group $i$.

At each time $t\in [0,T)$, the \emph{total size} of the population is given by
$$
N(t) = \sum_{i=1}^3 n_i(t),
$$
which is assumed to remain constant in time. Under this assumption, the distribution functions can be normalized with respect to $N_0 := N(0)$, hence we put
\begin{equation}\label{constant_size}
N(t) = N_0 = 1 \quad \mbox{for all}\quad t\in[0,T),
\end{equation}
so that $f_i$ defines the fraction of individuals belonging to the functional $i$-subsystem at each time $t$.

Additional macro-scale information is given by higher order moments:
\begin{equation}\label{moments}
\E_i^\nu[f_i](t) = \frac{1}{n_i(t)}\, \int_{D_u} u^\nu f_i(t, u)\, du.
\end{equation}
In particular, the first order moment $\mathbb{E}_i^1[f_i](t)$ will be simply denoted by $\mathbb{E}_i[f_i](t)$.

\subsection{\bf Micro-scale interactions}
 Let us consider the modeling of interactions at the microscopic scale, which  can modify the micro-state of the interacting pairs and/or promote transitions from one subsystem to another. Interactions between active particles involve the {\sl candidate} $h$-particle, namely belonging to functional subsystem $h$, and the {\sl field} $k$-particle, whose states are $u_*$ and $u^*$, respectively. After the interaction, the candidate $h$-particle can undergo a transition into the micro-state $u$ of the $i$th functional subsystem, namely into the state of the {\sl test} particle, which is representative of the whole system.
Another kind of interactions are those in which the candidate $h$-particle with state $u_*$ feels the mean value $\mathbb{E}_h$ within its functional subsystem \cite{[KNOP14]}. Therefore, individuals are subject to follow  a certain tendency, namely an attractive {\sl streaming effect}.

Theoretical tools of stochastic game theory are used to model the following terms related to the aforesaid interactions:
\begin{itemize}
\item  The encounter rate $\eta_{hk}(u_*,u^*)$ between a candidate $h$-particle with state $u_*$ and a field $k$-particle with state $u^*$.
\vspace{.1cm}
\item The interaction rate $\mu_{h}(u_*,\mathbb{E}_h)$ between a candidate $h$-particle, with state $u_*$, and the mean activity within its functional subsystem.
\vspace{.1cm}
\item  The transition probability density $\cB_{hk}^i(u_* \to u|u_*,u^*)$, which denotes  the
probability density that a  candidate $h$-particle with state $u_*$ ends up into the state $u$ of the test $i$-particle as a result of the interaction with a field $k$-particle with state $u^*$, satisfying that
\begin{equation}\label{density}
\sum_{i=1}^3 \int_{D_i} \mathcal{B}_{hk}^i (u_* \to u|u_*,  u^*)\, du = 1,
\end{equation}
for all type of inputs $(u_*,  u^*)$ and for all $h,k=1,2,3$.
\vspace{.1cm}
\item  The transition probability density $\mathcal{M}_{h}(u_*\to u | u_*, \mathbb{E}_h)$ denotes the probability that a candidate  $h$-particle with state $u_*$ ends up into the state $u$ after interacting with the mean activity value $\mathbb{E}_h$, satisfying that
\begin{equation}\label{density2}
\int_{D_h} \mathcal{M}_{h}(u_*\rightarrow u | u_*, \mathbb{E}_h) \, du = 1, \quad \hbox{for all} u_*\in D_h \quad \hbox{and} \quad \mathbb{E}_h.
\end{equation}
\end{itemize}

\subsection{\bf Derivation of a mathematical structure}
The balance of particles in the elementary volume of the space of micro-states leads to the following structure:
\begin{eqnarray}\label{structure}
&& \partial_t f_i (t,u) = J_i[{\bf f}](t,u)=
 \nonumber \\
 && \,   =  \sum_{h,k=1}^3 \int_{D_h} \int_{D_k} \eta_{hk}(u_*, u^*)
 \mathcal{B}_{hk}^i (u_* \to u| u_*,u^*) f_h(t, u_*) f_k(t, u^*) \, du_* \, du^* \nonumber \\
&& \quad - f_i(t,u) \sum_{k=1}^3 \,\int_{D_k} \, \eta_{ik}(u, u^*)\, f_k(t,u^*)\,  du^* \nonumber \\
&& \quad + \int_{D_i} \mu_{i}(u_*,\mathbb{E}_i) \mathcal{M}_{i}(u_*\rightarrow u | u_*, \mathbb{E}_i) f_i(t,u_*) du_* - \mu_{i}(u,\mathbb{E}_i) f_i(t,u)
\end{eqnarray}
for $i=1,2,3$, and where the square brackets are used to denote dependence on the whole
set of distribution functions $\f=\{f_i\}$, in the specific case it is on the mean value.

Before tackling technical issues it is worth discussing in some detail the proper role played by the  \emph{mathematical structure} (\ref{structure}) toward the derivation of particular models. As it is known \cite{[MAY04]}, the modeling of living systems cannot take advantage of field theories, like in the case of inert matter.  Therefore heuristic approaches are generally adopted, relying mainly on personal intuitions of the modelers, while more rigorous approaches can be developed by grounding models on the preliminary derivation of abstract mathematical structures consistent with the description presented in Section 2.

%%%%%%%%%%%%%%%%%%%%%%%%%%%%%%%%%%%%%%%%%%%%%%%%%%%%%
%%%%%%%%%%%%%%%%%%%%%%%%%%%%%%%%%%%%%%%%%%%%%%%%%%%%%
%%%%%%%%%%%%%%%%%%%%%%%%%%%%%%%%%%%%%%%%%%%%%%%%%%%%%

\section{Derivation of Models}
The derivation of the mathematical model is referred to the general structure given by Eq.(\ref{structure}) by particularizing the terms $\eta$, $\mu$, $\mathcal{B}$ and $\mathcal{M}$. The approach is limited only to the so-called  non-trivial interactions, which are  those that modify the micro-state of the interacting particles.

\subsection{Encounter rate}
Let us consider  the encounter rates $\eta_{hk}$ and $\mu_h$. The modeling approach is based on heuristic assumptions that let us quantify the frequency
of interactions, depending on the micro-states and distribution functions of the interacting particles.
Table 2 presents the encounter rates $\eta_{hk}$ for binary interactions between $h$-candidate and $k$-field particles,  while Table 3 refers to particles that interact with the mean activity value within their functional subsystem. $\eta_0$ and $\mu_0$ are positive constants in the expression of the encounter rates.

%%%%%%%%%%%%%%%%%%%%%%%%%%%%%%%%%%%%%%%%%%%%%%%%%%%%%%%%%%%%%%%%%%%%%%%%%%%%%%%%%%%%%%%%%%%%%%
\begin{table}[h!] \label{table}
\begin{center}
\begin{tabular}{ccc}
\hline
\textbf{Interaction} & \textbf{Qualitative description} &  \textbf{Encounter rate} \, $\eta$  \\
\hline
& Closer social states &  \\						
\framebox{\footnotesize 1}  $\leftrightarrow$ \textcircled{\footnotesize 1} & tend to interact&$\eta_{11}(u_*,u^*)=\eta^{0} \left(1-|u_*-u^*|\right)$  \\
  & more frequently &\\						
\hline
& Experienced lawbreakers& \\						
\framebox{\footnotesize 2}  $\leftrightarrow$ \textcircled{\footnotesize 2}  & are more expected to& $\eta_{22}(u_*,u^*) = \eta^{0} (u_*+u^*)$\\
& expose themselves & \\						
\hline
& & \\												
\framebox{\footnotesize 2}  $\leftrightarrow$ \textcircled{\footnotesize 3}  & Experienced detectives &$\eta_{23}(u_*,u^*) = \eta^{0} \big((1-u_*) + u^*\big)$ 		    \\
 & are more likely to {\em hunt} &\\	
% \hline\\
\framebox{\footnotesize 3}  $\leftrightarrow$ \textcircled{\footnotesize 2} & less experienced criminals &$\eta_{32}(u_*,u^*) = \eta^{0} \big(u_* + (1-u^*)\big)$ 		    \\
\hline
\end{tabular}
\end{center}
\vskip.2cm
\caption{Non-trivial interactions between a $h$-candidate particle (represented by a square) with state $u_*$ and a $k$-field particle (represented by a circle) with state $u^*$.}
\end{table}

%%%%%%%%%%%%%%%%%%%%%%%%%%%%%%%%%%%%%%%%%%%%%%%%%%%%%%%%%%%%%%%%%%%%%%%%%%%%%%%%%%%%%%%%%%%%%%

%%%%%%%%%%%%%%%%%%%%%%%%%%%%%%%%%%%%%%%%%%%%%%%%%%%%%%%%%%%%%%%%%%%%%%%%%%%%%%%%%%%%%%%%%%%%%%
\begin{table}[h!] \label{table}
\begin{center}
\begin{tabular}{ccc}
\hline
\textbf{Interaction} & \textbf{Qualitative description} &  \textbf{Encounter rate} \, $\mu$  \\
\hline
& Criminals  interact with &  \\						
\framebox{\footnotesize 2}  $\leftrightarrow$ $\mathbb{E}_2$ &  with the mean value through & $\mu_{2}(u_*,\mathbb{E}_2)= \mu^{0}|u_* - \mathbb{E}_2|$  \\
  &  the mean-micro state distance  &\\						
\hline
& Detectives interact with &\\						
\framebox{\footnotesize 3}  $\leftrightarrow$ $\mathbb{E}_3$  & with the mean value through  & $\mu_{3}(u_*,\mathbb{E}_3) = \mu^{0}|u_*-\mathbb{E}_3|$ \\	
& the mean micro-state distance &\\						
\hline
\end{tabular}
\end{center}
\vskip.2cm
\caption{Non-trivial interactions between a $h$-candidate particle (represented by a square) with activity $u_*$ and the mean activity value $\mathbb{E}_h$.}
\end{table}

%%%%%%%%%%%%%%%%%%%%%%%%%%%%%%%%%%%%%%%%%%%%%%%%%%%%%%%%%%%%%%%%%%%%%%%%%%%%%%%%%%%%%%%%%%%%%%

\subsection{Transition probability densities}

The modeling of the transition probability densities $\cB_{hk}^i(u_* \to u|u_*,u^*)$ is performed under the simplified assumption that the output of the interaction is a delta function over the most probable value. This choice needs dealing with some technical features related to the modeling of interactions as Eq.~(\ref{structure}) requires that the output of the interaction  belongs to the open interval $(0,1)$ and the distribution functions have compact support on this interval. The parameters of the model  take values in $[0,1)$. Table 4 defines their meaning.

%%%%%%%%%%%%%%%%%%%%%%%%%%%%%%%%%%%%%%%%%%%%%%%%%%%%%%%%%%%%%%%%%%%%%%%%%%%%%%%%%%%%%%%%%%%%%%
\begin{table}[h!] \label{table}
\begin{center}
\begin{tabular}{ll}
\hline
\textbf{Parameters} &  \\
\hline
$\alpha_T \quad$  Susceptibility of citizens to become criminals \\
\hline
$\alpha_B \quad$  Susceptibility of criminals to reach back the state of normal citizens \\
\hline
$\beta \quad$  Learning dynamics among criminals  \\
 \hline
 $\gamma \quad$  Motivation/efficacy of security forces  to catch criminals  \\
\hline
$\lambda \quad$  Learning dynamics among detectives  \\
\hline
\end{tabular}
\end{center}
\vskip.2cm
\caption{Parameters involved in the table of games.}
\end{table}

%%%%%%%%%%%%%%%%%%%%%%%%%%%%%%%%%%%%%%%%%%%%%%%%%%%%%%%%%%%%%%%%%%%%%%%%%%%%%%%%%%%%%%%%%%%%%%

Bearing this in mind,  the so-called {\sl table of games}, that gives the probability distributions of candidate's payoffs conditioned to the states of the interacting individuals, is proposed according to the following assumptions:

\begin{itemize}

\item The social structure of the population $i=1$ is assumed to be fixed,  namely, the time interval is sufficiently short
 that the wealth distribution is  constant in time.

\vskip.1cm \item  Citizens are susceptible to become criminals, motivated by their wealth state. More in detail, a candidate citizen with activity $u_*$ interacting with a richer one with activity $u^*>u_*$ can become a criminal, mutating into functional subsystem $2$ with a very low criminal ability $u = \varepsilon \approx 0$. In particular it is assumed that the transition probability increases with decreasing wealth:
\begin{equation}\label{B11}
\begin{cases}
 \cB_{11}^2(u_* \to u |u_*,u^*) =\dfrac1\varepsilon \alpha_T \, (1-u_*)u^*  \chi_{[0,\varepsilon)}(u),  \\
 \\
 \cB_{11}^1(u_* \to u |u_*,u^*) = (1-\alpha_T \, (1-u_*)u^*) \delta_{u_*}(u),
\end{cases}
\end{equation}
where $\chi_{[0,\varepsilon)}$ denotes the indicator function for the interval $[0,\varepsilon)$.

\vskip.1cm \item Criminals interact among themselves resulting in a dynamics by which less experienced criminals  mimic the more experienced ones, moreover
also interaction with less experienced lawbreakers increases the level of criminality
\begin{equation}\label{B22}
\cB_{22}^2(u_* \to u |u_*,u^*) = \delta_{u_*+\beta(1-u_*)u^*}(u),
\end{equation}
where $0 \leq \beta < 1$ .

\vskip.1cm \item Detectives of functional subsystem $h=3$ chase criminals of functional subsystem
$k=2$ and the latter are constrained to step back decreasing their activity value
as the price to be paid for being caught. At the same time, detectives gain experience from a well-done job increasing their activity. Due to this action
criminals are induced to return to the state of normal citizens with probability which increases with decreasing values of their level of criminality and increasing values of skill of detectives:
\begin{equation}\label{B32}
\cB_{32}^3(u_* \to u |u_*,u^*) = \delta_{u_*+\gamma u^*(1-u_*)}(u),
\end{equation}
and
\begin{equation}\label{B33}
\begin{cases}
\cB_{23}^1(u_* \to u |u_*,u^*) = \dfrac1\varepsilon \alpha_B (1 - u_*)u^* \chi_{[0,\varepsilon)}(u),\\
\\
\cB_{23}^2(u_* \to u |u_*,u^*) = (1 - \alpha_B (1 - u_*)u^*) \, \delta_{u_*-\gamma u^*u_*}(u).
\end{cases}
\end{equation}
\end{itemize}

The dynamics of interactions is visualized in Figures 1--3, where black and gray bullets correspond, respectively, to pre-interaction and post-interaction states.
\begin{figure} \label{interactions1}
 \includegraphics[width=0.6\textwidth]{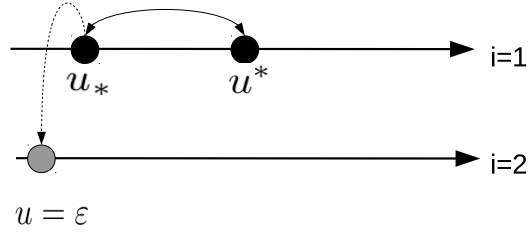}
 \vskip-.5cm
\caption{Interaction between citizens may end up, in the mutation of one of the interacting individuals, into the functional subsystem of criminals.}
\end{figure}
\begin{figure} \label{interactions2}
\includegraphics[width=0.6\textwidth]{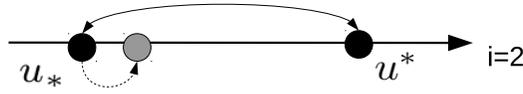} \\
 \vskip-1cm
\caption{Progression to higher values of criminality.}
\end{figure}
\begin{figure} \label{interactions3}
  \includegraphics[width=0.7\textwidth]{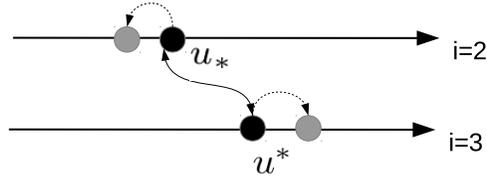} \\
   \vskip-1cm
\caption{Interaction between detectives and criminals may cause a reduction of the criminal's activity and an increase of the detective's activity.}
\end{figure}

\subsection{Modeling the stream effect}

Functional subsystems of criminals and detectives are subject to interactions
with their respective mean activity values.
The dynamics is similar to that of Eq.(\ref{B22}), in which only those who are less experienced
than the mean tend to learn and move towards it
\begin{equation}\label{M22}
\mathcal{M}_{2}(u_* \to u |u_*,\mathbb{E}_2) = \delta_{\beta u_*+(1-\beta)\mathbb{E}_2}(u).
\end{equation}

Analogously, also detectives show a trend toward the mean value
\begin{equation}\label{M33}
\mathcal{M}_{3}(u_* \to u |u_*,\mathbb{E}_3) = \delta_{\lambda u_*+(1-\lambda)\mathbb{E}_3}(u),
\end{equation}
where $0\le \lambda < 1$ is related to the tendency of detectives to approach
the mean security level of the forces.

\subsection{Derivation of the model}
The mathematical structure (\ref{structure}) can be specified for each distribution function
$f_i$ by means of the tables of games given by Eqs.(\ref{B11})-(\ref{M33}) and the interaction
rates proposed in Tables 2 and 3. Accordingly, the evolution equations read:
\begin{eqnarray}\label{m1}
&& \partial_t f_1(t,u)  =
 -\alpha_T (1-u)f_1(t,u)\int_0^1\eta_{11}(u,u^*)u^*f_1(t,u^*)du^*\nonumber \\
 && \, +\frac1\varepsilon\alpha_B\chi_{[0,\varepsilon)}(u)\int_0^1\int_0^1\eta_{23}(u_*,u^*)(1-u_*)u^*f_2(t,u_*)f_3(t,u^*)du_*du^*,
\end{eqnarray}
\begin{eqnarray}\label{m2}
&&\partial_t f_2(t,u) =
\frac1\varepsilon \alpha_T \chi_{[0,\varepsilon)}(u)\int_0^1\int_0^1\eta_{11}(u_*,u^*)(1-u_*)u^*f_1(t,u_*)f_1(t,u^*)du_*du^*\nonumber \\
&& \quad +\int_0^1\chi_{[\beta u^*,1]}(u)\frac1{1-\beta u^*}\eta_{22}\left(\frac{u-\beta u^*}{1-\beta u^*},u^*\right)f_2\left(t,\frac{u-\beta u^*}{1-\beta u^*}\right)f_2(t,u^*)du^*\nonumber \\
&& \quad +\int_0^1\chi_{[0,1-\gamma u^*]}(u)\frac1{1-\gamma u^*}\eta_{23}\left(\frac{u}{1-\gamma u^*},u^*\right)\left[1-\alpha_B\left(1-\frac{u}{1-\gamma u^*}\right)u^*\right]\nonumber \\
&&\quad\qquad \cdot f_2\left(t,\frac{u}{1-\gamma u^*}\right) f_3(t,u^*)du^*\nonumber  \\
&&  \quad-f_2(t,u)\sum_{k=2}^3\int_0^1\eta_{2k}(u,u^*)f_k(t,u^*)du^*\nonumber \\
&&  \quad+\frac1\beta \chi_{[(1-\beta)\mathbb{E}_2,\beta+(1-\beta)\mathbb{E}_2]}(u)\mu_2\left(\frac{u-(1-\beta) \mathbb{E}_2}\beta,\mathbb{E}_2\right)f_2\left(t,\frac{u-(1-\beta)\mathbb{E}_2}\beta\right)\nonumber \\
&& \quad-\mu_2(u,\mathbb{E}_2)f_2(t,u),
\end{eqnarray}
and
\begin{eqnarray}\label{m3}
&&\partial_t f_3(t,u) =
 \int_0^1\chi_{[\gamma u^*,1]}(u)\frac1{1-\gamma u^*}\eta_{32}\left(\frac{u-\gamma u^*}{1-\gamma u^*},u^*\right)\nonumber \\
&&\quad\cdot f_3\left(t,\frac{u-\gamma u^*}{1-\gamma u^*}\right)f_2(t,u^*)du^* -f_3(t,u)\int_0^1\eta_{32}(u,u^*)f_2(t,u^*)du^*\nonumber \\
&&\quad+\frac1\lambda\chi_{[(1-\lambda)\mathbb{E}_3,\lambda+(1-\lambda)\mathbb{E}_3]}(u)\mu_3\left(\frac{u-(1-\lambda) \mathbb{E}_3}\lambda,\mathbb{E}_3\right)f_3\left(t,\frac{u-(1-\lambda)\mathbb{E}_3}\lambda\right)\nonumber \\
&&\quad-\mu_3(u,\mathbb{E}_3)f_3(t,u).
\end{eqnarray}

%%%%%%%%%%%%%%%%%%%%%%%%%%%%%%%%%%%%%%%%%%%%%%%%%%%%%
%%%%%%%%%%%%%%%%%%%%%%%%%%%%%%%%%%%%%%%%%%%%%%%%%%%%%
%%%%%%%%%%%%%%%%%%%%%%%%%%%%%%%%%%%%%%%%%%%%%%%%%%%%%

\section{Simulations and Critical Analysis}
The mathematical structure proposed in Eq. (3.7) has generated, as we have seen, a model stated in terms of a system of ordinary differential equations. Coupling this system to the initial conditions, the statement of the initial value problem is as follows:
\begin{equation}\label{ivp}
\begin{cases}\partial_t f_i(t,u)=J_i[\f](t,u),&\\
f_i(0,u)=f_i^0(u)&
\end{cases}\quad\mbox{for } i=1,2,3,
\end{equation}
where the operators $J_i[\f]$ have been defined in Eq. \eqref{structure} and in the specific model Eqs. \eqref{m1}--\eqref{m3}, and $f_i^0\in L^1(0,1)$, with $f_i^0\ge 0$ a.e. in $[0,1]$ for $i=1,2,3$. Let us now consider the space $\boldsymbol{X}=[L^1(0,1)]^3$, endowed with the norm
$$
\|\boldsymbol{\psi}\|=\sum_{i=1}^3\|\psi_i\|_1=\sum_{i=1}^3\int_0^1|\psi_i(u)|du,$$
and let
$$
X_+=\left\{\psi\in L^1(0,1)\,:\, \psi\ge0 \mbox{ a.e. in } [0,1]\right\}\mbox{ and }\boldsymbol{X}_+=\left\{\boldsymbol{\psi}\in \boldsymbol{X}\,:\, \psi_i\in X_+, \, i=1,2,3\right\}
$$
denote the positive cones of $L^1(0,1)$ and $\boldsymbol{X}$, respectively.

A natural solution space for problem \eqref{ivp} is $C^1([0,T),\boldsymbol{X})$, $0<T\le\infty$, that is the space of $[L^1(0,1)]^3$--functions $\f=\f(t)$ of class $C^1([0,T))$.

We shall denote by $\boldsymbol{\f^0}=(f_1^0,f_2^0,f_3^0)$. Due to the normalization (3.3), we assume that $\|\f^0\|=1$. This implies immediately that if $\f$ is a non-negative solution of \eqref{ivp} (i.e. $\f\in C^1([0,T),\boldsymbol{X}_+)$), then $\|\f(t,\cdot)\|=1$ for all $t\in [0,T)$, by Eqs. \eqref{density}--\eqref{density2}.

The following classical existence and uniqueness theorem can be stated:
\begin{theorem}\label{teu}
Let $\boldsymbol{\f_0}$ be in $\boldsymbol{X}_+$, and suppose that there exist two positive constants $C_\eta,\, C_\mu$ such that
$$\sup_{(u_*,u^*)\in[0,1]^2}\eta_{hk}(u_*,u^*)\le C_\eta, \quad\sup_{u_*\in[0,1]}\sup_{\psi_i\in X_+}\mu_i(u_*, \mathbb E_i[\psi_i])\le C_\mu,$$
%\sup_{(u_*,u)\in[0,1]^2}\sup_{\phi_i\in X_+}\mathcal M_i(u_*,u,\mathbb E_i[\phi_i])\le C_{\mathcal M}\end{gathered}$$
for all $h,\,k,\,i=1,2,3$. Let $\mathcal M_i=\mathcal M_i(u_*,u,\mathbb E_i[\psi_i])$ and $\mu_i=\mu_i(u_*,\mathbb E_i[\psi_i])$ be Lipschitz continuous with respect to $\psi_i$ for all $i=1,2,3$.
Then \eqref{ivp} admits a unique solution $\f\in C^1([0,\infty),\boldsymbol X_+)$.
\end{theorem}

This result follows from an application of the Banach fixed point theorem, where the Lipschitz property of the right hand side of the first equation in \eqref{ivp} leads to local existence and uniqueness, while positivity is shown by taking the exponential form of the differential equation.
Then, a continuation argument leads to existence and uniqueness of solutions for large times. The technical details of the proof are given in Appendix. The interested reader is also addressed to \cite{[ADFLB12],[CS13],[PS14]} for proofs referred to equations with similar structure and analytic properties.

\vspace{.1cm}
Simulations have to be selected to put in evidence emerging behaviors of interest to test the predictive ability of the model. Therefore, some specific case studies are selected with the aim  to investigate the relationship between the social structure and the levels of criminality in a society. In particular, it is important to understand how the prosperity of a society and the social differences between individuals impact on the rise of criminal acts.  The focus is limited not only to the influence of the mean wealth value, but also to the shape of the wealth distribution at equal mean wealth value.

 These two topics are treated in the next two subsections, while the importance of the number of effective security agents to fight against criminality is studied in the third subsection. Then, although the development of an exhaustive sensitivity analysis of all parameters is not treated here, a detailed study of the role of two of them is proposed in  in the fourth subsection, namely focusing on the susceptibility of citizens to become criminals $\alpha_T$ and the efficacy of security forces $\gamma$, in order to test their importance in the prevention of crime. Finally,  Table 5 summarizes the result of all simulations.

%%%%%%%%%%%%%%%%%%%%%%%%%%%%%%%%%%%%%%%%%%%%%%%%%%%%%%%%%%%%%%%%%%%%%%%%%%%%%%%%%%%%%%%%%%%%%%%%%%%%%%

\subsection{Case 1: Dynamics for different mean  wealth values}

Simulations aim at depicting the time evolution of the number of criminals, starting from the ideal situation of $n_2(t=0) = 0$, for different values of the mean wealth of citizens. With that purpose we took a variety of initial conditions $f_1(0,u)$, all of which had a small rich cluster and a larger low-middle class cluster, as shown in Fig.~4(a), with mean wealth taking values $\mathbb{E}_1 = 0.25$, $0.35$, $0.44$ and $0.53$.  The solution is computed for large times and Fig.~4(b) represents the final distribution of criminals $f_2$. Only two curves are shown in this figure for the sake of clarity. Figure~4(c) shows the evolution of the size of the population of criminals over time, $n_2(t)$, for these different values of $\mathbb{E}_1$. In all cases, simulations were developed for the following values of parameters: $\alpha_T = 0.01$, $\alpha_B = 0.1$, $\beta = 0.1$, $\gamma = 0.25$ and $\lambda = 0.9$. These figures show that a poor society leads to high levels of crime. This trend holds for a broad variety of parameters.

\begin{figure}[h!]\label{CS1_fig1} %Figures generated in the folder CS_mean_wealth
\begin{tabular}{cc}
\includegraphics[width=0.43 \textwidth]{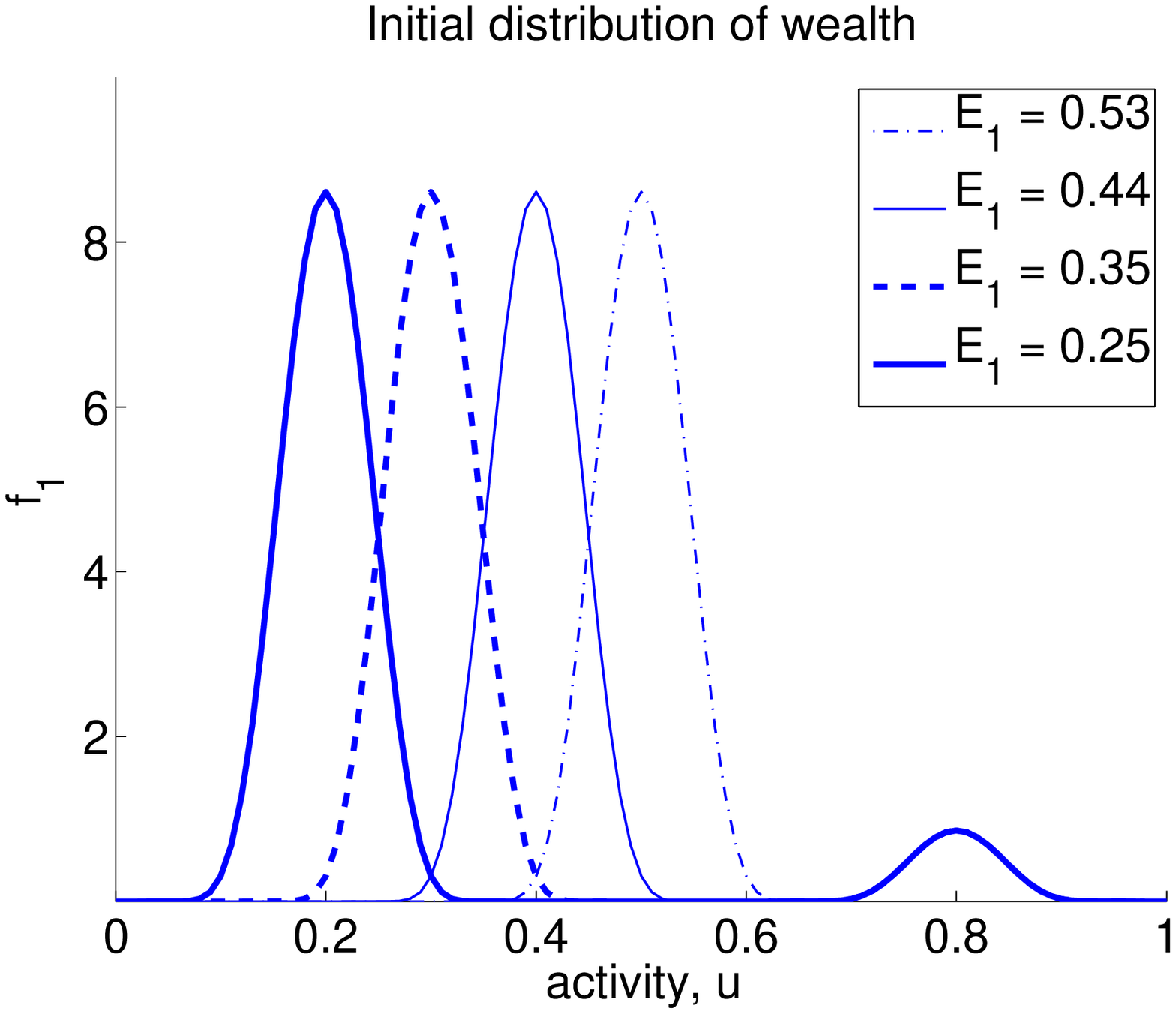} &
\includegraphics[width=0.43 \textwidth]{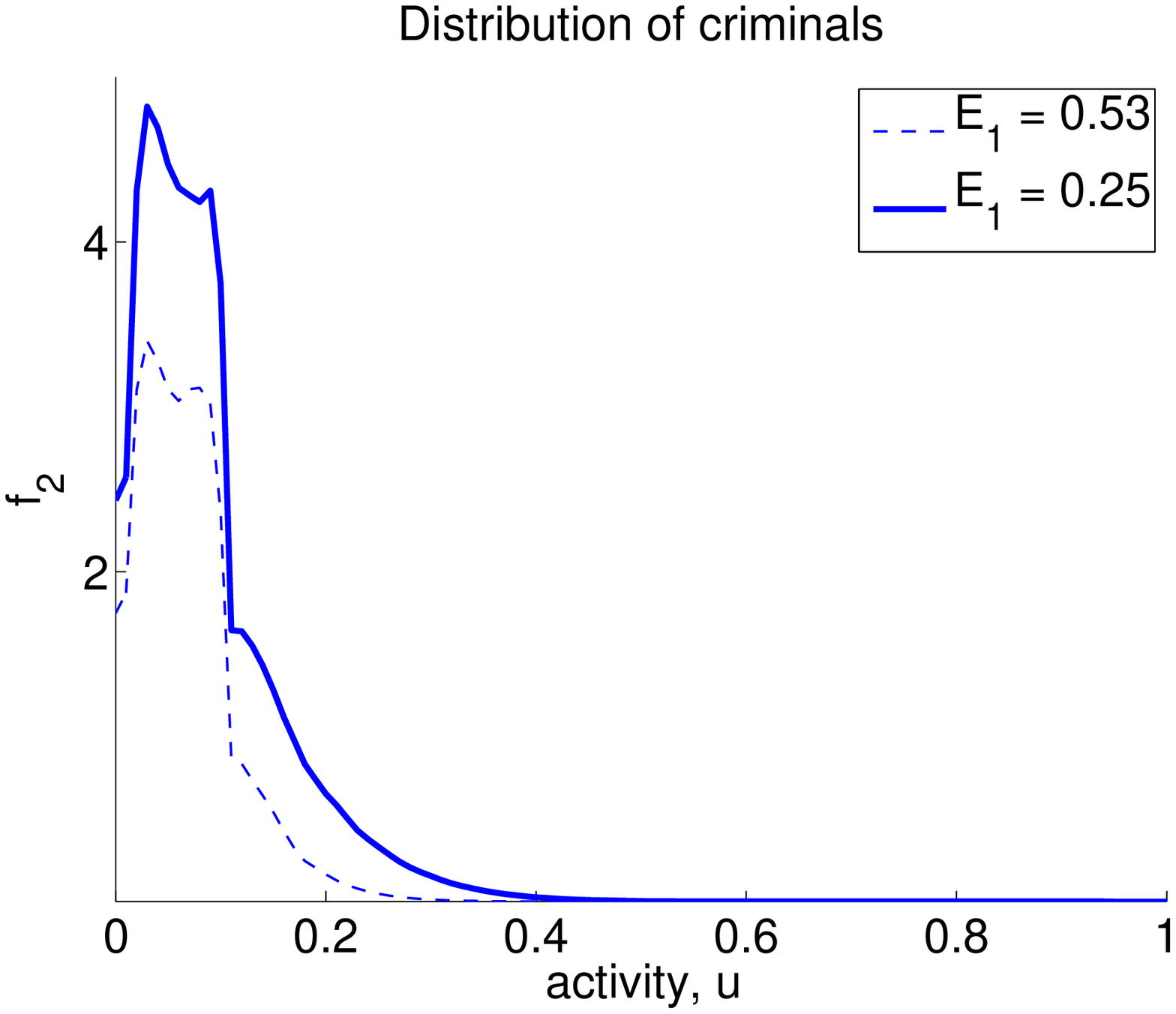} \\ %in the folder CS_mean_wealth_bis
(a) & (b)
\end{tabular}
\begin{center}
\includegraphics[width=0.43 \textwidth]{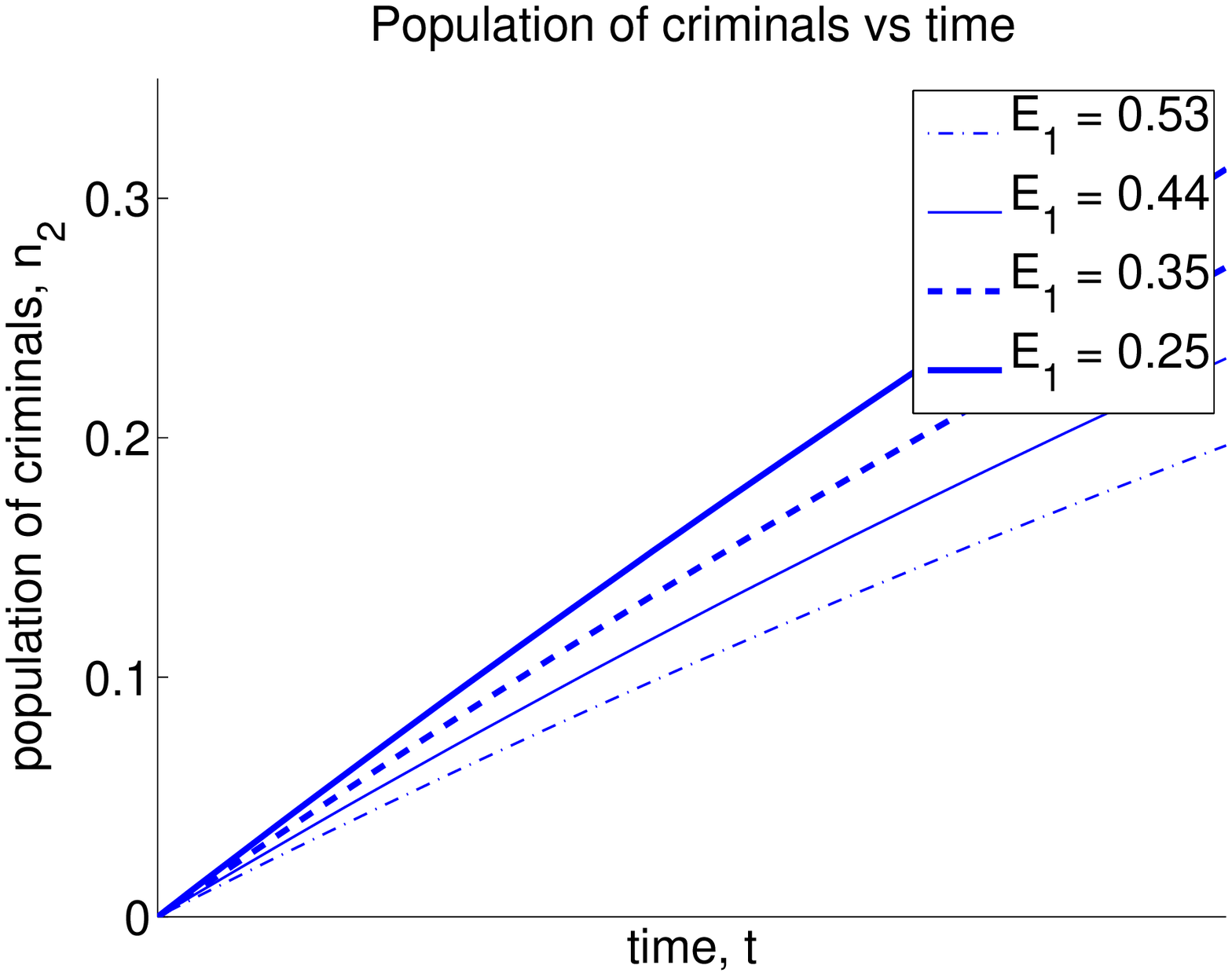} \\ %in the folder CS_mean_wealth
(c)
\end{center}
\caption{(a) Initial wealth distributions corresponding to different mean wealth values. All of them consist in a fixed small rich cluster and a large poorer cluster centered in different points of the activity domain. (b) Large time distribution of criminals for two of the selected mean wealth values. (c) Evolution of the size of functional subsystem $2$, $n_2(t)$, for different values of $\mathbb{E}_1$.}
\end{figure}

%%%%%%%%%%%%%%%%%%%%%%%%%%%%%%%%%%%%%%%%%%%%%%%%%%%%%%%%%%%%%%%%%%%%%%%%%%%%%%%%%%%%%%%%%%%%%%%%%%%%%%
\subsection{Case 2: Dynamics for different shapes of  wealth distribution}

Simulations are developed corresponding to fixed values of the mean wealth, specifically two values are selected, low ($\mathbb{E}_1 = 0.2$) and high ($\mathbb{E}_1 = 0.6$), while two different shapes are considered for each case corresponding to higher and lower concentrations of wealth in the middle class, as depicted in Fig.~5(a) and 5(c). In this case we take a non-zero initial condition for the number of criminals, considering that $n_2(0)/n_1(0) = 0.05$ (corresponding to a society where the initial number of criminals is 5\% of the number of citizens), and we define the quantity
$$
\varphi(t) = \frac{n_2(t)-n_2(0)}{n_2(0)}\cdot 10^2,
$$
as a measure of the relative percentage change in the population of criminals. Simulations were developed for the set of parameters: $\alpha_T = 0.0001$, $\alpha_B = 0.15$, $\beta = 0.1$, $\gamma = 0.15$ and $\lambda = 0.9$, Figs.~5(b) and 5(d) report the evolution of $\varphi$. We can observe that a poorer society produces a growth in the number of criminals, that is still more accentuated for unequal wealth distributions. The model is capable to produce the opposite behavior for a richer society, giving a reduction in the number of criminals for the same choice of parameters.

\begin{figure}[h!] %Figures generated in the folder CS_different_shapes
\begin{tabular}{cc}
\includegraphics[width=0.43 \textwidth]{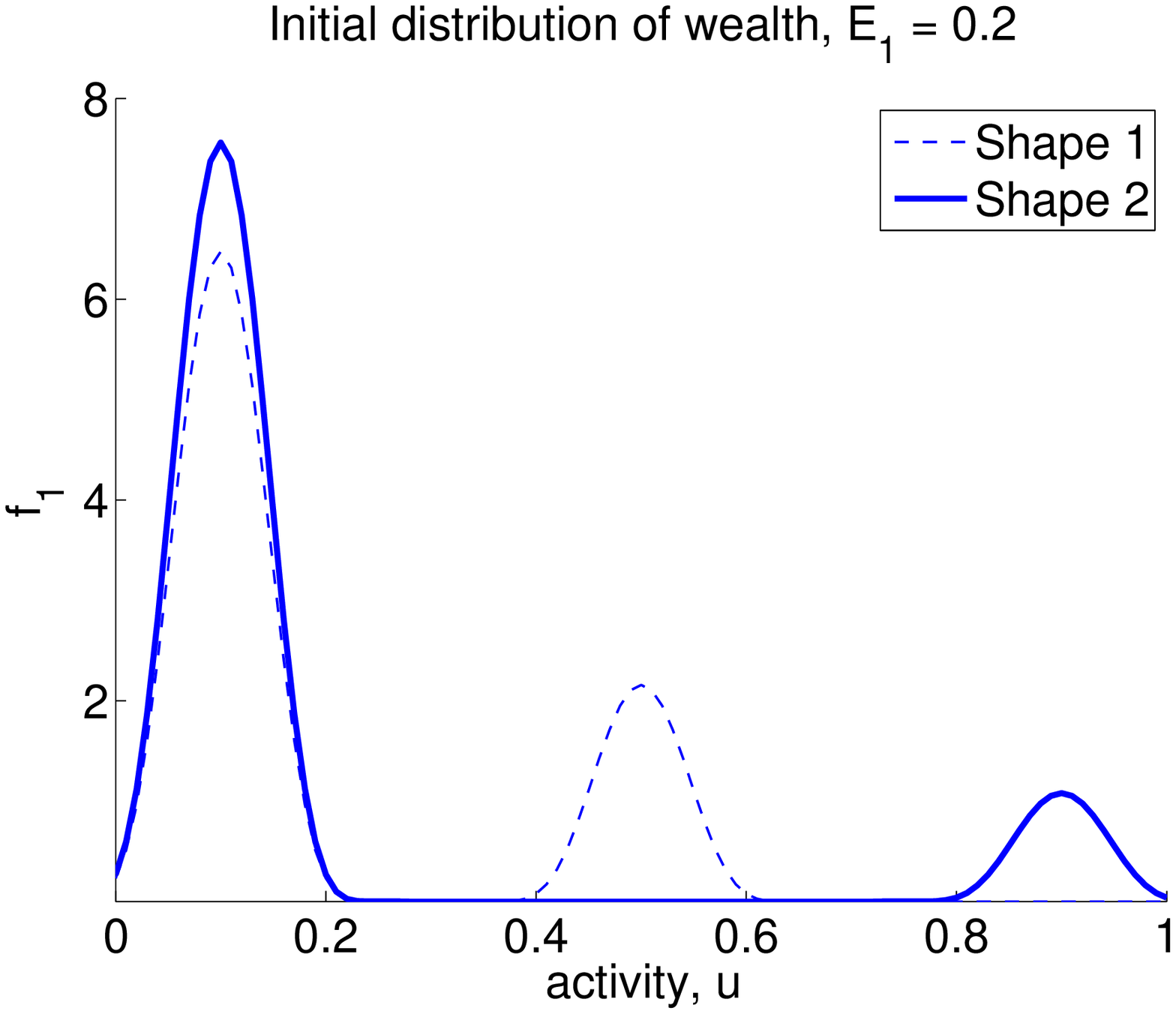} & %generated with file main.m
\includegraphics[width=0.43 \textwidth]{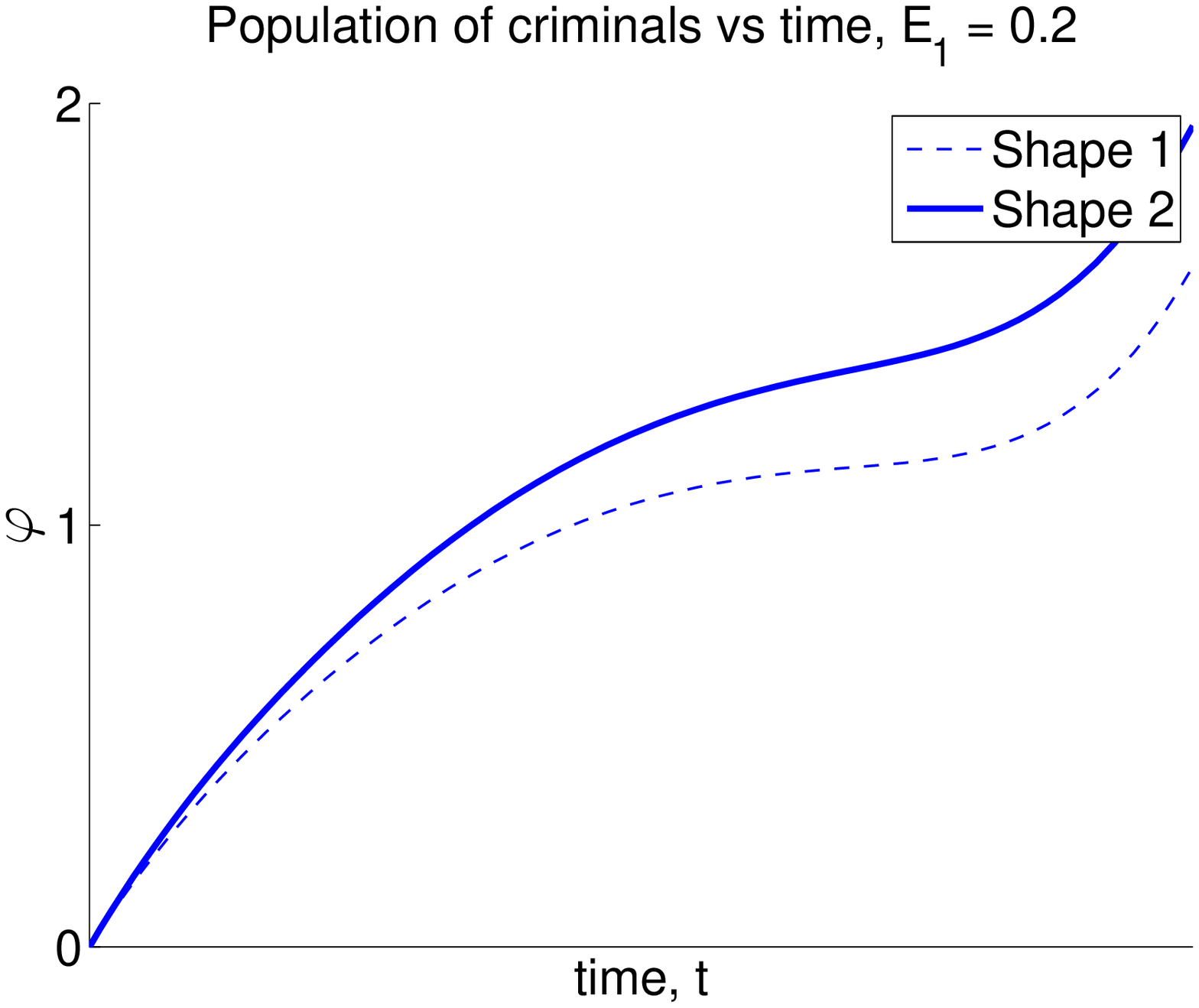} \\
(a) & (b) \\
\includegraphics[width=0.43 \textwidth]{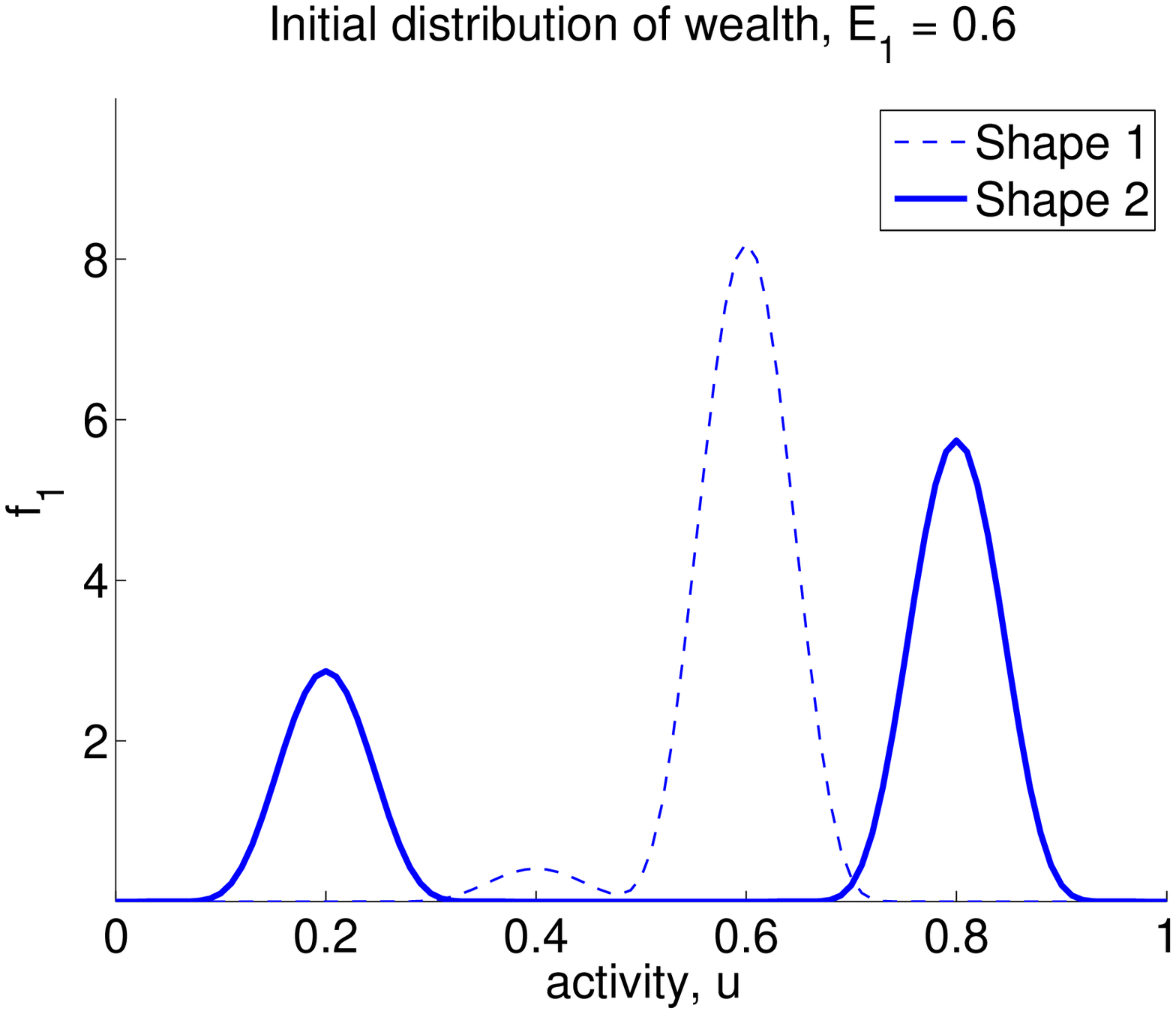} & %generated with file main_rich.m
\includegraphics[width=0.43 \textwidth]{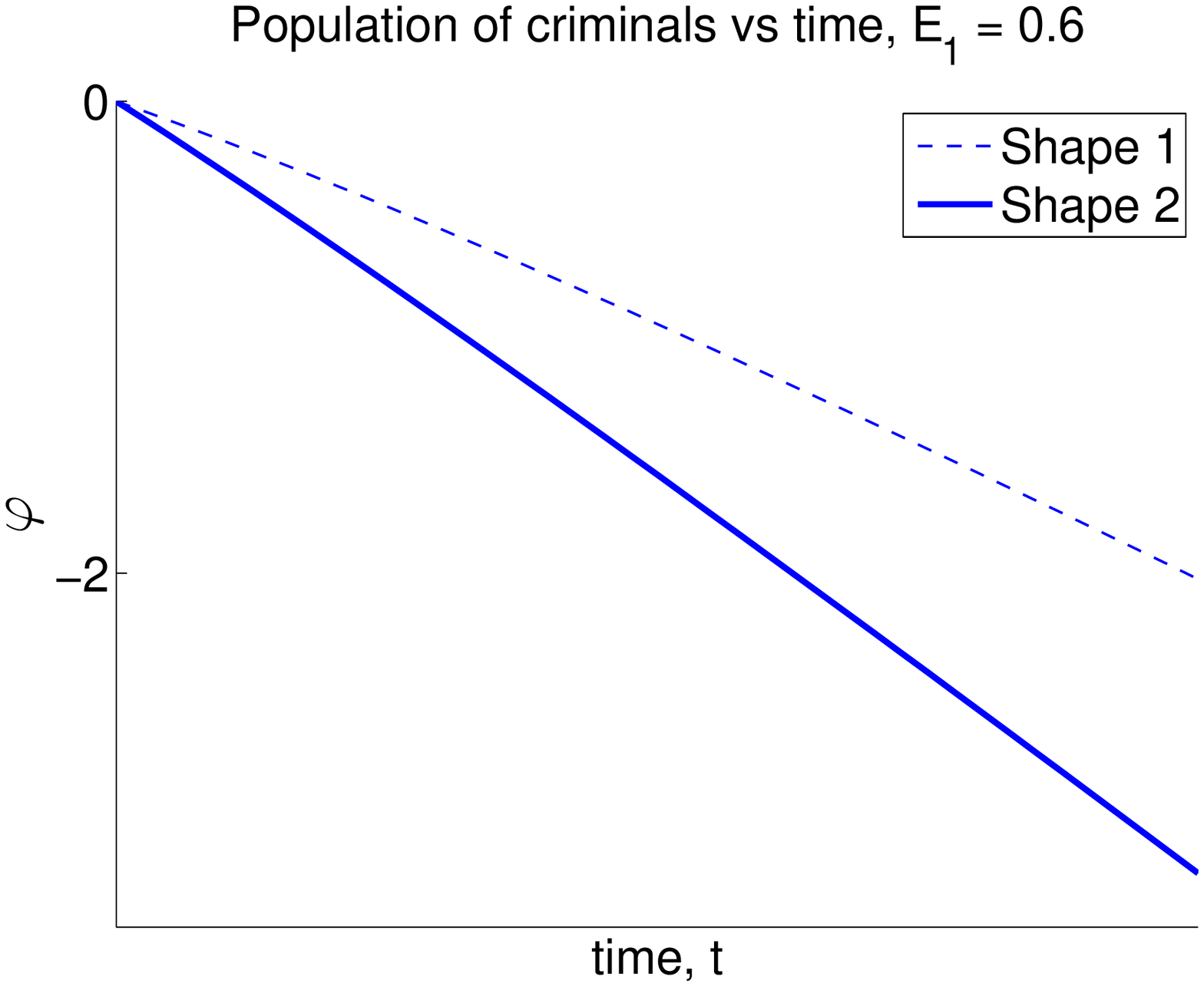} \\
(c) & (d) \\
\end{tabular}
\caption{(a) Two wealth distributions for a poor society with $\mathbb{E}_1 = 0.2$ lead to different curves for the relative change of the population of criminals (b), where the most unequal distribution generates a larger increase. Analogously, (c) shows two wealth distributions for a rich society with $\mathbb{E}_1 = 0.6$ that generate, for the same set of model parameters, a reduction in the number of criminals (d).}
\end{figure}

%%%%%%%%%%%%%%%%%%%%%%%%%%%%%%%%%%%%%%%%%%%%%%%%%%%%%%%%%%%%%%%%%%%%%%%%%%%%%%%%%%%%%%%%%%%%%%%%%%%%%%
\subsection{Case 3: On the role of the number of detectives}

Another interesting topic consists in investigating the influence of the number of detectives in the development of crime. Taking into account the worldwide distribution of police agents per country, the rates have a median of 303.3 officers per 100,000 people and a mean of 341.8 officers \cite{[HHM10]}. Of course, these numbers depend on the social and structural differences between countries. Let us show the evolution of the number of criminals, taking a fixed initial mean wealth $\mathbb{E}_1 = 0.433$ and a fixed ratio $n_2(0)/n_1(0) = 0.1$, for different values of $n_3(0)/n_1(0)$ (in particular we consider two cases: 1110 and 330 detectives per 100,000 citizens). In all cases the initial distribution of detectives $f_3(0,\cdot)$ is a Gaussian--type function with mean value $0.5$.  Figure 6(a) shows the initial distribution of criminals $f_2(0,\cdot)$ and Fig.~6(b) shows the large time distribution for different values of $n_3(0)/n_1(0)$.  We take the set of parameters: $\alpha_T = 0.0001$, $\alpha_B = 0.15$, $\beta = 0.05$, $\gamma = 0.01$ and $\lambda = 0.9$.  In Figs.~6(c) and 6(d) we show the evolution of the macroscopic quantities $\varphi(t)$ and $\mathbb{E}_2(t)$. The results show that for this selection of parameters the expected number cited in \cite{[HHM10]} keeps the number of criminals under control. Larger squads contribute to the reduction in the number of criminals as well as in the mean criminal ability.

\begin{figure}[h!]\label{CS_number_of_detectives} %Figures generated in the folder CS_number_of_detectives_bis
\begin{tabular}{cc}
\includegraphics[width=0.43 \textwidth]{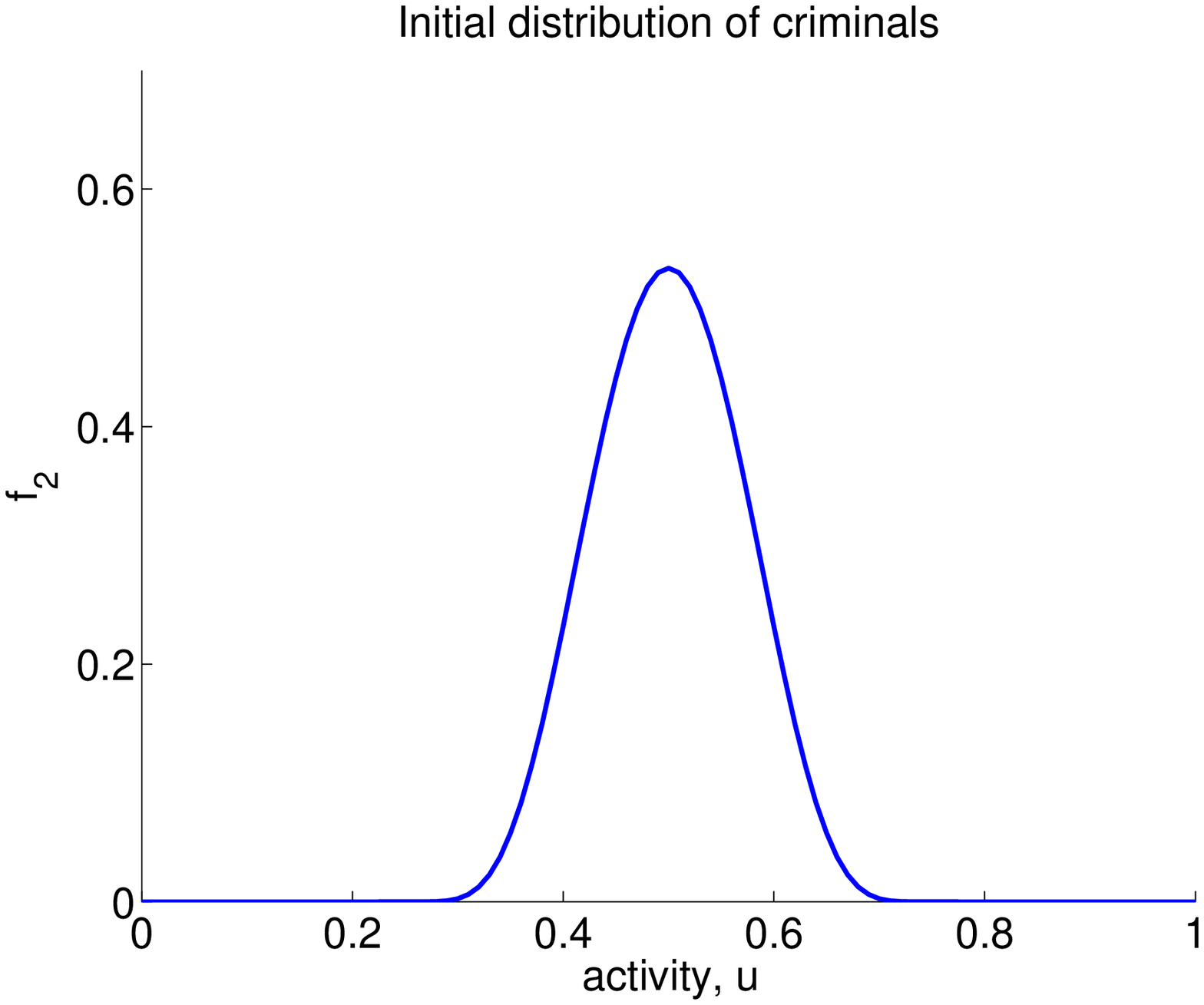} &
\includegraphics[width=0.43 \textwidth]{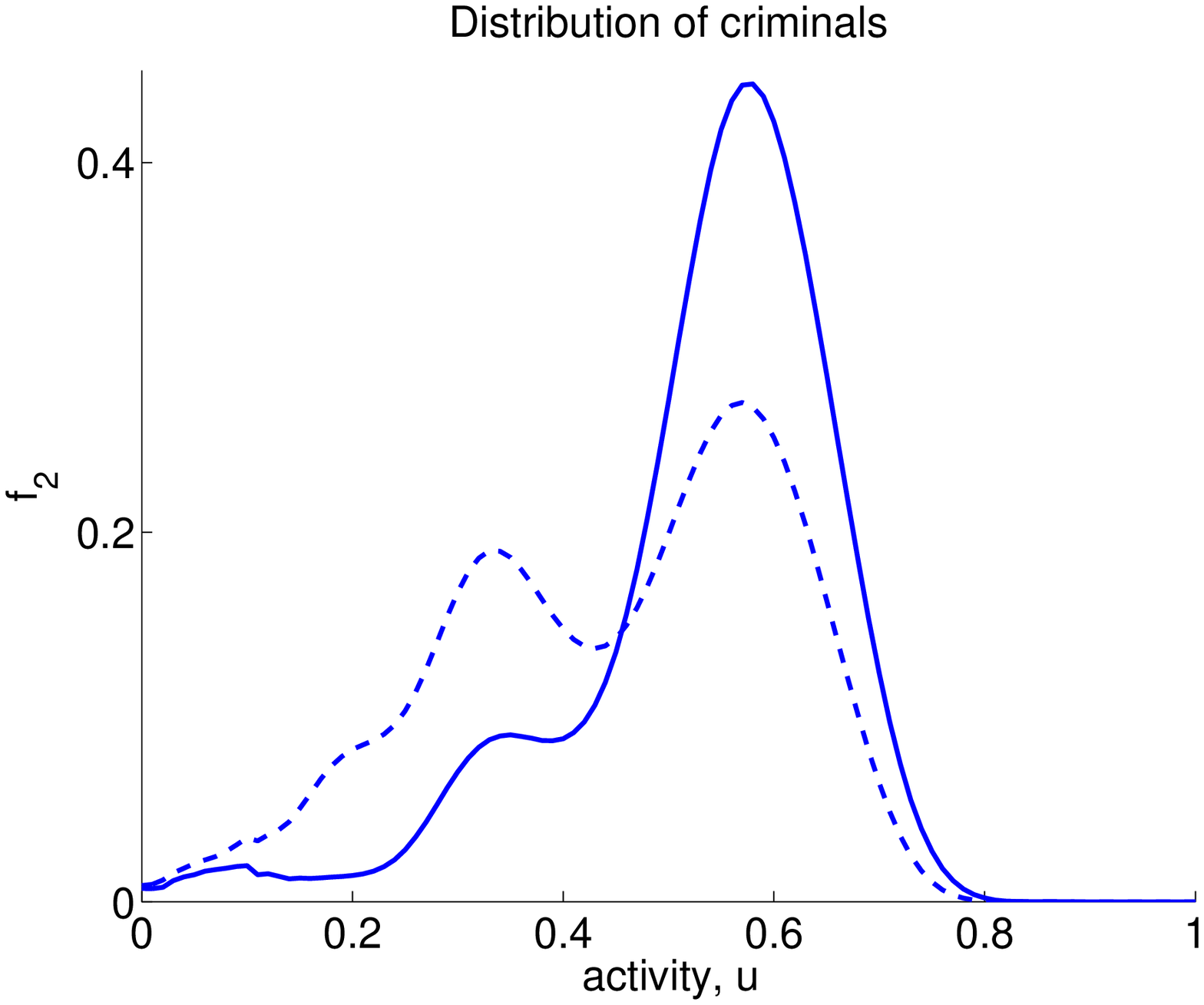} \\
(a) & (b) \\
\includegraphics[width=0.43 \textwidth]{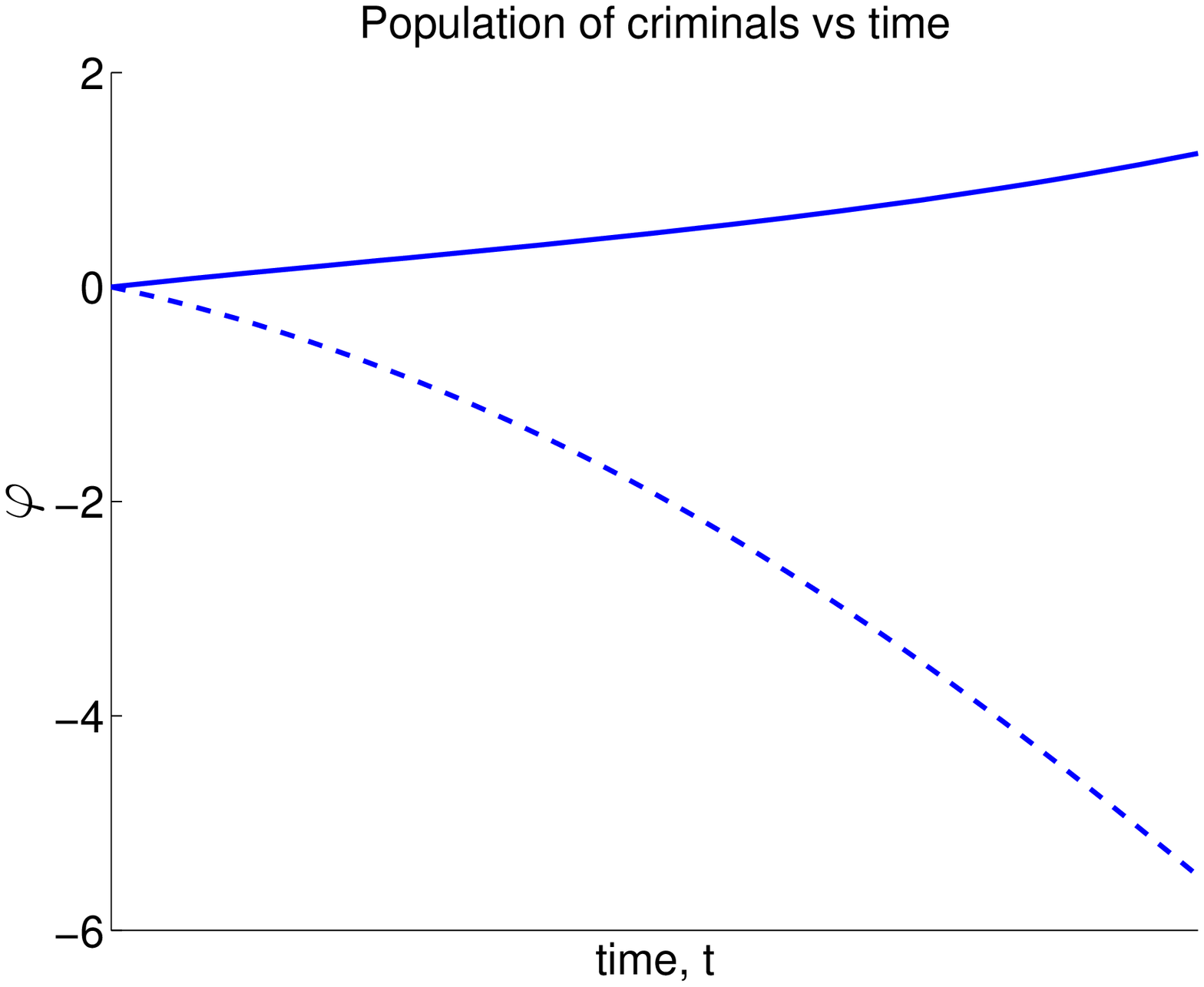} &
\includegraphics[width=0.43 \textwidth]{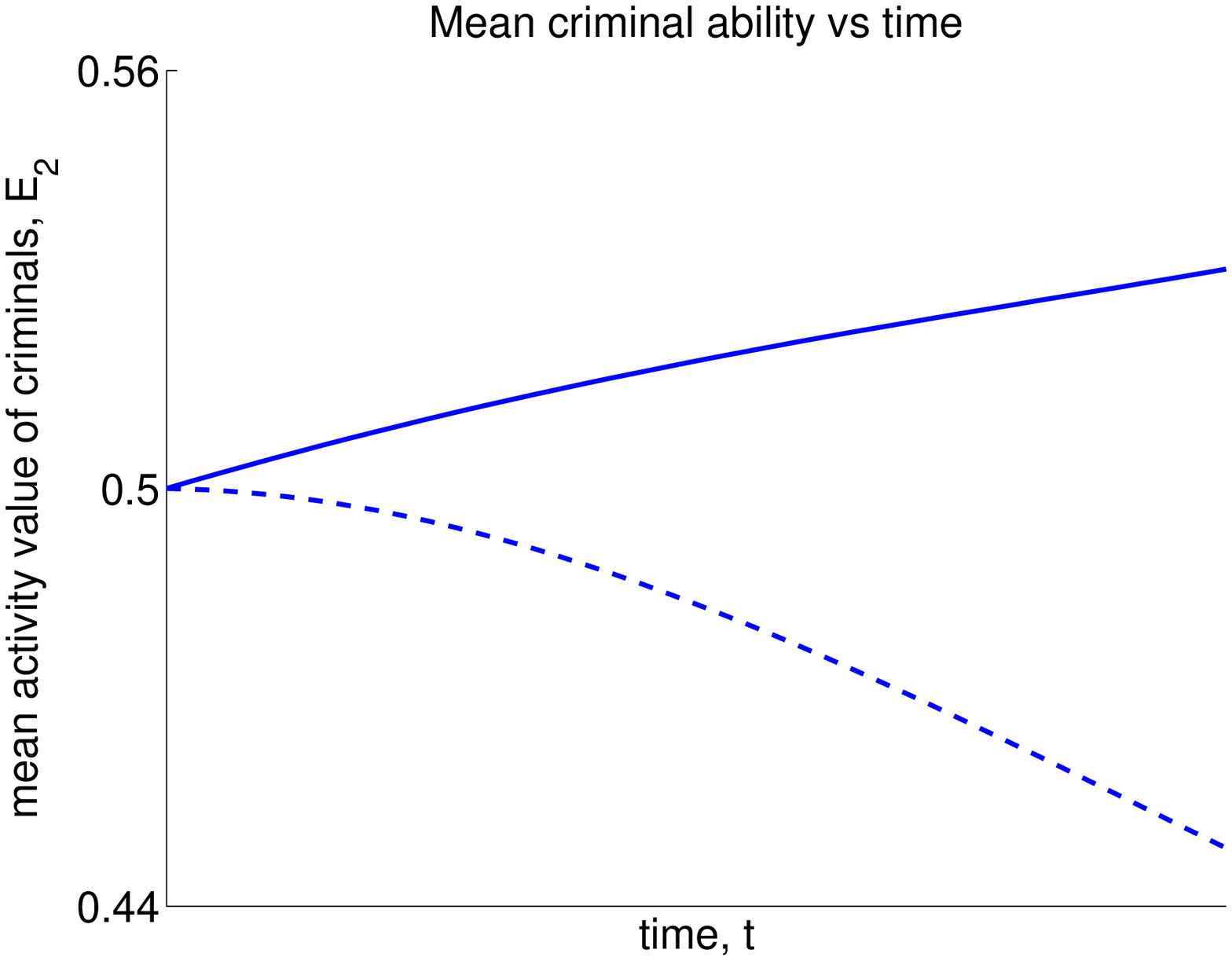} \\
(c) & (d)
\end{tabular}
\caption{(a) Initial distribution of criminals. (b) Large time distribution of criminals, (c) evolution of $\varphi(t)$ and (d) evolution of the mean criminal ability, $\mathbb{E}_2(t)$, for different number of security agents per citizen. Continuous lines correspond to $n_3(0)/n_1(0)=3.3\cdot 10^{-3}$ and dashed lines correspond to $n_3(0)/n_1(0)=11.1\cdot 10^{-3}$.}
\end{figure}

%%%%%%%%%%%%%%%%%%%%%%%%%%%%%%%%%%%%%%%%%%%%%%%%%%%%%%%%%%%%%%%%%%%%%%%%%%%%%%%%%%%%%%%%%%%%%%%%%%%%%%
\subsection{Case 4: Prevention of crime, is it better to improve the well-being of the society or to strengthen  security forces?}

Consider a population initially distributed with $n_2(0)/n_1(0) = 0.05$ and with  500 detectives per 100,000 citizens and let us study the evolution of $\varphi(t)$ with variation of the parameters $\alpha_T$ and $\gamma$. The initial distribution of detectives $f_3(0,\cdot)$ is a Gaussian--type function with mean value $0.5$. Figure 7(a) shows the time dynamics of $\varphi$  for different values of $\alpha_T$,  and for $\alpha_B = 0.05$, $\beta = 0.1$, $\gamma = 0.5$ and $\lambda = 0.9$.  Fig.~7(b) shows the same dynamics for  $\alpha_T = 0.0002$, and different values of $\gamma$. This figure confirms the empirical evidence that an  effective action to fight  crime consists in pursuing actions that contribute to reduce $\alpha_T$ (education, employment, etc) and to improve the quality of citizens \cite{[ORM05]}.
\begin{figure}[h!]\label{CS_parameter_sensitivity} %Figures generated in the folder CS_parameter_sensibility
\begin{tabular}{cc}
\includegraphics[width=0.43 \textwidth]{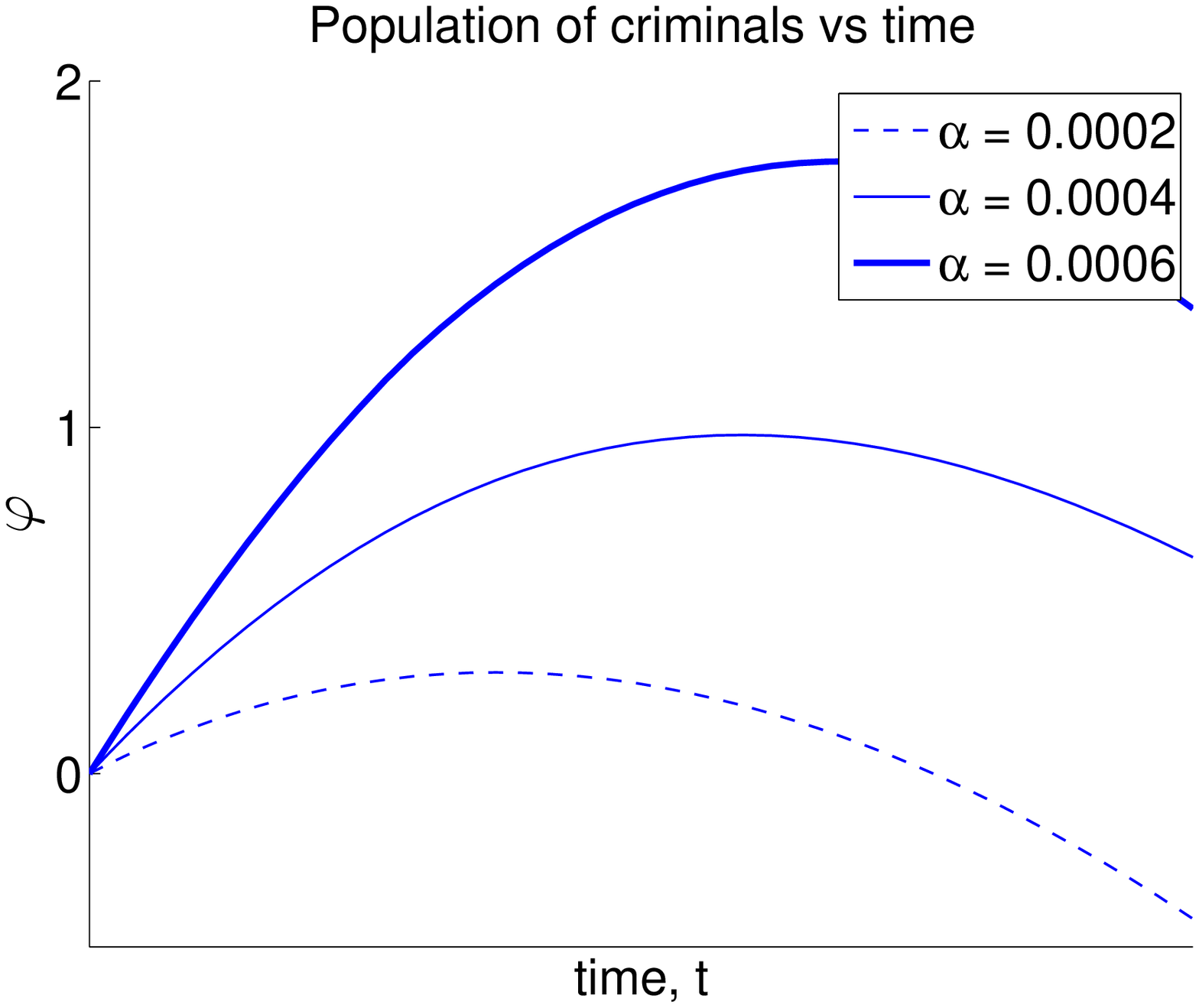} &
\includegraphics[width=0.43 \textwidth]{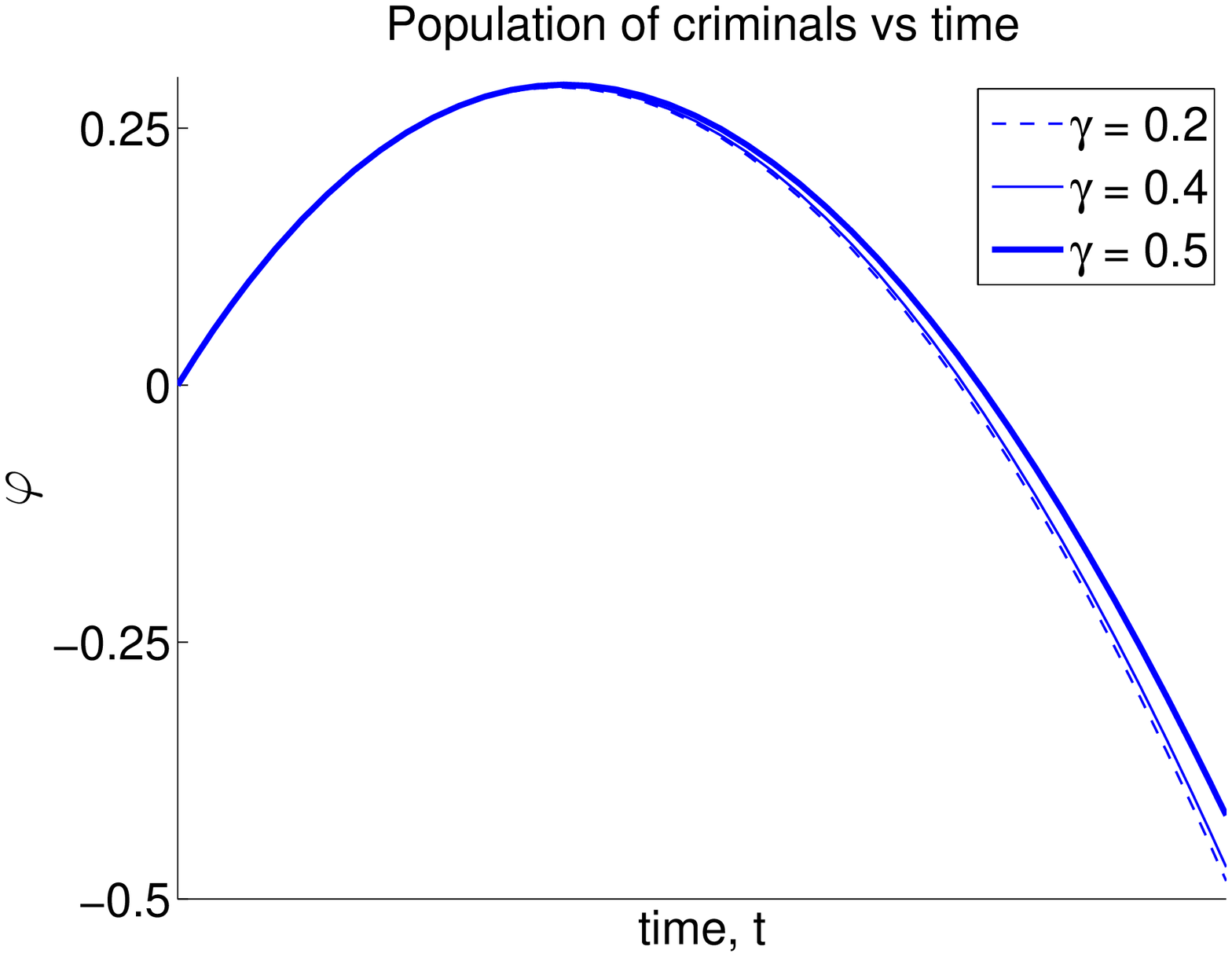} \\
(a) & (b)
\end{tabular}
\caption{Time evolution of  $\varphi(t)$ for functional subsystem $2$, for different values of (a) $\alpha_T$ and (b) $\gamma$.}
\end{figure}

%%%%%%%%%%%%%%%%%%%%%%%%%%%%%%%%%%%%%%%%%%%%%%%%%%%%%%%%%%%%%%%%%%%%%%%%%%%%%%%%%%%%%%%%%%%%%%%%%%%%%%
\subsection{Critical analysis}

Simulations performed in the previous subsections tested the predictive ability of the model
regarding to the raise of criminality in a society according basically to the wealth
distribution within it (cases 1 and 2), the importance of the number of detectives
fighting against crime (case 3) and the advantages of developing policies that guarantee
a better well-being (case 4).

\begin{center}
\begin{table}[h!] \label{table}
{\bf Case 1 - Role of the mean wealth} \\
\vspace*{.2cm}
\begin{tabular}{|c|c|c|}
\hline
 &  &  Increasing number of criminals \\
Decreasing mean wealth of the society & $\Longrightarrow$\qquad &  \\
 &  &  Increasing criminal ability \\
\hline
\end{tabular}\\
\vspace*{.5cm}
{\bf Case 2 - Role of the shape of wealth distribution}\\
\vspace*{.2cm}
\begin{tabular}{|c|c|c|c|}
\hline
 & Equal distribution & $\Longrightarrow$ & Slow growth in the number \\
 &  &  & of criminals  \\
  Poor society & & & \\
 & Unequal distribution & $\Longrightarrow$ & Fast growth in the number \\
  & & & of criminals \\
 \hline
 & Equal distribution & $\Longrightarrow$ & Fast decrease in the number\\
  &  &  &  of criminals   \\
 Rich society &  &  &  \\
 & Unequal distribution & $\Longrightarrow$ & Slow decrease in the number \\
 & & &  of criminals \\
\hline
\end{tabular}\\
\vspace*{.5cm}
{\bf Case 3 - Role of the number of detectives}\\
\vspace*{.2cm}
\begin{tabular}{|c|c|c|}
\hline
 &  &   \\
Large number of detectives & $\Longrightarrow$ & Decreasing number of criminals  \\
 &  &  \\
\hline
\end{tabular}\\
\vspace*{.5cm}
{\bf Case 4 - Role of parameters $\alpha_T$ and $\gamma$}\\
\vspace*{.2cm}
\begin{tabular}{|c|c|c|}
\hline
 & $\Longrightarrow$\qquad  & Number of criminals under control   \\
Low susceptibility & & \\
to criminality  & $\Longrightarrow$\qquad & Criminal ability under control \\
\hline
& $\Longrightarrow$\qquad  &  Decreasing number of criminals (with little  \\
Increasing ability  & & sensitivity)\\
of detectives & $\Longrightarrow$\qquad & Decreasing criminal ability \\
 \hline
\end{tabular}\\
\vspace*{.5cm}
\caption{Summary of the simulations results.}
\end{table}
\end{center}

The different emerging behaviors, corresponding to the aforementioned computational study, are presented and summarized in Table 5.
Although the computational study cannot be considered exhaustive, still some useful indications are delivered. The most important one is,
according to the authors' bias, the need of a detailed analysis of the interplay of different dynamics. In this specific case, the welfare
policy and that of the fight against crime. Previous studies, namely \cite{[BHT13]} - focused on the dynamics of support and opposition to regimes -
and \cite{[DL14]} - focused on the growth or decay of global wealth - have shown that the interaction of two dynamics can lead to non-expected
outputs, in some cases even to non-predictable events. Presumably, the investigation is worth to be developed even in this specific social problem studied
in this paper.

%%%%%%%%%%%%%%%%%%%%%%%%%%%%%%%%%%%%%%%%%%%%%%%%%%%%%
%%%%%%%%%%%%%%%%%%%%%%%%%%%%%%%%%%%%%%%%%%%%%%%%%%%%%
%%%%%%%%%%%%%%%%%%%%%%%%%%%%%%%%%%%%%%%%%%%%%%%%%%%%%

\section{Looking Forward}\label{complete}

The contents of this paper suggest various perspectives for further research activity in the field focusing on simulations and model validation, analytic problems, and understanding the role of space dynamics. Some of topics is presented at an introductory level in the next  subsections according to a selection based on the authors' bias. For each of them some hints for future research activity will be given.

\subsection{Critical size of individuals.}
The  survival of functional subsystems might, in some cases, depend on their size. If the size  of a
functional subsystem falls below a critical value $N_c$, then interactions reduce to transfer individuals from the
original one towards an aggregation to another one or the dynamics is modified.  For instance, for detectives, their action is negligible if the critical size goes below a certain level needed for an efficient action.

%%%%%%%%%%%%%%%%%%%%%%%%%%%%%%%%%%%%%%%%%

\subsection{Topological domains of interactions}\label{7.2}
Interactions depend also on the domain of the activity variable, within which each particle has the ability to perceive a sufficient amount of signals and develop consequently a strategy \cite{[BCC08],[BS12]}.  According to \cite{[BS12]}, interactions occur in a domain $\Omega \subseteq D_u$, while sufficient information is achieved if  a number $n_c$ of field particles is involved. Integration of $f_i$ over the activity variable in a domain  $\Omega = [u - s_m[f;n_c], u + s_M[f;n_c]]$, which can be called  the topological domain of interaction, can lead to compute $s_m$ and $s_M$, where $s_m, s_M > 0$. However, the solution is unique only in some special cases. For instance, when $u$ is a scalar defined over the whole real axis, and the sensitivity is symmetric with respect to $u$.
On the other hand, if $u$ is defined in a bounded domain or the sensitivity is not symmetric, additional assumptions are required.

 The set $\Omega$ is a part of the effective interaction domain, which is the intersection of the theoretical maximum domain of activity of an individual with the support of $f_i(t, \cdot)$. The topological domain is concerned with the possible activities of individuals, defined by their effective domain.  The choice of the topological domain depends in this way on the neighborhood of the activity but also on more preferred activities of the individuals, although not necessarily measured in the Euclidean distance.

\subsection{Dynamic of sub-domains. Local and global interactions}
This issue is also related to the previous Subsection \ref{7.2}. Let us first define the concept of sub-domains we will deal with and
assume that each functional population has the same global interests. One can imagine a scenario in which any population is subdivided in  different groups that may develop different or complementary strategies among them or to cooperate or have a consensus with individuals or groups of other populations.
For example, consider that the criminal population is divided into two factions with common interests, although they can compete (or collaborate) with resources based on different strategies in connection with some other individuals or groups (police or ordinary citizens) that may have confluent strategy towards a common benefit, at least temporarily. These relationships can bring this subgroup into more social positions or just to castle on criminal ones.
Each of these subgroups  defines a sub-domain of the population of criminals, that shares the main features of the whole population but has some specific rules in relation with the other group. Emergent processes are also a consequence of the interactions of these sub-domains of the different populations.

On the other hand, the dynamics of each sub-domain might be defined in terms of the characteristic curves associated with the activity variable. The flux across these curves must be characterized through the potential generated by the activity, that involves the sub-domains of the different populations. This is a challenging issue that needs  more precise and detailed definitions which will be discussed in a forthcoming paper.

%%%%%%%%%%%%%%%
%%%%%%%%%%%%%%%%%%%%%%%%%%%%%%%%%%%%%%%%%%%%%%%%%%

\subsection{On the role of external actions}\label{7.4}

The model presented in the  Section 4 does not include an analysis of the role of external actions, which can be applied to reduce criminality either by repressive (or persuasion)  actions, or by improving the expertise of detectives. In principle, also an educative action can be addressed to citizens to improve their ability to protect their  goods. A simple model consists in assuming that a macro-scale action, say ${\mathcal U}_i(t,u)$ depending on time and activity variables is applied to the functional subsystems corresponding to the second and third functional subsystem.

The modeling of this action is achieved by inserting it in the transport term, which is linear if the action depends only on $t$ and $u$ , while it is nonlinear when it depends also on the distribution functions,  ${\mathcal U}_i[f_i](t,u)$. Of course these actions have a cost, which have to be related to the benefit of the criminality reduction. Classically, if this aspect is introduced a control and optimization problem can be developed.

%%%%%%%%%%%%%%%%%%%%%%%%%%%%%%%%%%%%%%%%%%%%%%%%%%%%%

\subsection{On the role of space dynamics}

Models and  simulations proposed in the paper refer to a population homogeneously distributed in space. A variety of real world applications suggests to extend the study to populations distributed in networks \cite{[KNOP14]}, where interaction between nodes can, in some cases, play an important role in the overall dynamics, for instance inducing migration phenomena  \cite{[KNOP13]}. Also the distribution over space is of practical interest to understand the real localization of criminality and their level of danger in specific areas of the territory. The classical problem consists in deriving macroscopic PDEs models from the underlying description at the microscopic scale. The pioneer ideas of paper \cite{[ODA88]}, developed in \cite{[BBNS07]} for hyperbolic scaling for multicellular systems with internal structures, need additional nontrivial studies to take into account the heterogeneous features of the territory.

%%%%%%%%%%%%%%%%%%%%%%%%%%%%%%%%%%%%%%%%%%%%%%%%%%%%%
%%%%%%%%%%%%%%%%%%%%%%%%%%%%%%%%%%%%%%%%%%%%%%%%%%%%%
%%%%%%%%%%%%%%%%%%%%%%%%%%%%%%%%%%%%%%%%%%%%%%%%%%%%%
\appendix

\section{Proof of Theorem \ref{teu}}
\begin{proof}
Consider the Banach space $(\mathscr C^{T_1},\|\cdot\|_\infty)$, where
$$\mathscr C^{T_1}=C([0,T_1],\boldsymbol{X})\quad\mbox{ and }\quad\|\f\|_\infty=\max_{t\in[0,T_1]}\|\f(t,\cdot)\|$$
and $T_1>0$ will be specified later. Put $a>1$ and
$$\mathscr C_+^{T_1}=\left\{\f\in \mathscr C^{T_1}\,:\, f_i(0,u)=f_i^0(u),\, f_i(t,u)\ge0,\, \|\f\|_\infty\le a \,\, \mbox{ for all } t,\,u,\,i\right\}.$$
Clearly, $\mathscr C_+^{T_1}$ is a non-empty, closed subset of the Banach space $\mathscr C^{T_1}$.

Let us now introduce the operator $\mathscr S: \mathscr C_+^{T_1}\to\mathscr C^{T_1}$, defined for all $\f\in\mathscr C_+^{T_1}$ as
$$
\begin{aligned}(\mathscr S&\f)_i(t,u)=e^{-Ct}\bigg\{f_i^0(u)+\int_0^t e^{Cs}\Big[\sum_{h,k=1}^3\int_0^1\int_0^1\eta_{hk}(u_*, u^*)
 \mathcal{B}_{hk}^i (u_* \to u| u_*,u^*) \\
&\times f_h(s, u_*) f_k(s, u^*) \, du_* \, du^* - f_i(s,u)\left( \sum_{k=1}^3 \,\int_0^1 \, \eta_{ik}(u, u^*)\, f_k(s,u^*)\,  du^*-a \tilde{C}_\eta\right) \\
& + \int_0^1 \mu_{i}(u_*,\mathbb{E}_i) \mathcal{M}_{i}(u_*\rightarrow u | u_*, \mathbb{E}_i) f_i(s,u_*) du_* -  f_i(s,u)(\mu_{i}(u,\mathbb{E}_i)-C_\mu)\Big]ds \bigg\}\end{aligned}
$$
for all $(t,u)\in[0,T_1]\times[0,1]$ and $i=1,2,3$, where $C=a\tilde{C}_\eta+C_\mu$ and $\tilde{C}_\eta=\max\{1/5,C_\eta\}$.

Let $L_\mu$ and $L_{\mathcal M}$ denote the Lipschitz constants of $\mu_i(u_*,\mathbb E_i[\psi_i])$ and $\mathcal M_i(u_*,u,\mathbb E_i[\psi_i])$ with respect to $\psi_i$, respectively, and take $T_1>0$ such that
\begin{equation}\label{T1}T_1<\frac1C\min\left\{\ln\frac{3a}{1+2a},\ln\frac{5C+2aL_\mu+aL_{\mathcal M}C_\mu}{5C+2aL_\mu+aL_{\mathcal M}C_\mu-1}\right\}.\end{equation}

The local existence and uniqueness of the solution follows from an application of the Banach fixed point theorem. Indeed, it is easy to check that if $\f$ is a solution of \eqref{ivp} in $[0,T_1]\times[0,1]$ then $\f$ is a fixed point of $\mathscr S$ and vice versa. Furthermore, $\mathscr S(\mathscr C^{T_1}_+)\subset \mathscr C^{T_1}_+$, since for all $\f\in\mathscr C^{T_1}_+$ it results that $\mathscr S\f(0,\cdot)=\f_0$, $(\mathscr S\f)_i\ge 0$ for $i=1,2,3$,  and for all $t\in[0,T_1]$
$$
\begin{aligned}\sum_{i=1}^3\int_0^1(\mathscr S\f)_i(t,u)du\le& e^{-Ct}\bigg\{\|\f^0\|+\int_0^t e^{Cs}\Big[2\tilde{C}_\eta\|\f(s,\cdot)\|^2+a\tilde{C}_\eta\|\f(s,\cdot)\|\\
&+3C_\mu\|\f(s,\cdot)\|\Big]ds\bigg\}\le e^{-Ct}+3a(1-e^{-Ct})\\
\le &1+3a(1-e^{-CT_1})\le a,\end{aligned}$$
where \eqref{T1} is taken into account.

Now, proceeding as in the proof of \cite[Theorem 3.1]{[PS14]}, we obtain,  for all $\f,\boldsymbol{g}\in \mathscr C_+^{T_1}$, the following inequality:
$$
\|\mathscr S \f-\mathscr S \boldsymbol{g}\|_\infty =\max_{t\in[0,T_1]}\sum_{i=1}^3\int_0^1|\mathscr S\f(t,u)-\mathscr S\boldsymbol{g}(t,u)|du\le k\|\f-\boldsymbol{g}\|_\infty,
$$
where $k=(1-e^{-CT_1})(5C+2aL_\mu+aL_{\mathcal M}C_\mu)<1$, by \eqref{T1}.

Therefore, the operator $\mathscr S:\mathscr C_+^{T_1}\to\mathscr C_+^{T_1}$ is a contraction, and by the Banach fixed point theorem there exists a unique solution $\f$ of \eqref{ivp} defined in $[0,T_1]\times[0,1]$. By \eqref{density}--\eqref{density2}, $\|\f(T_1,\cdot)\|=1$, then, by iterating the reasoning as in the proof of \cite[Theorem 3.5]{[CS13]}, $\f$ can be extended uniquely to a global solution of \eqref{ivp}.\end{proof}

 \end{document}